\documentstyle[aps,prc,preprint]{revtex}
\begin{document}
\draft
\title{
Measurement of the Electric and Magnetic Polarizabilities of the Proton}
\author{B. E. MacGibbon, G. Garino \cite{Garino}, M. A. Lucas \cite{Lucas},
and  A. M. Nathan}
\address{Department of Physics, University of
Illinois at Urbana-Champaign, \\
Urbana, Illinois 61801}
\author{G. Feldman}
\address{Saskatchewan Accelerator Laboratory, University of Saskatchewan, \\
Saskatoon, Saskatchewan Canada S7N 0W0}
\author{B. Dolbilkin}
\address{Institute for Nuclear Research, Russian Academy of Sciences \\
Moscow 117312 Russia}
\date{\today}
\maketitle
\begin{abstract}
The Compton scattering cross section on the proton has been
 measured at laboratory
angles of 90$^\circ$ and 135$^\circ$ using tagged photons in the energy
range 70--100 MeV and
simultaneously using untagged photons in the range 100--148~MeV.
With the aid of dispersion relations, these cross sections were
used to extract the electric and magnetic polarizabilities,
$\bar{\alpha}$ and $\bar{\beta}$ respectively, of the proton.
We find
$$\bar{\alpha}+\bar{\beta}  =  ( 15.0 \pm 2.9 \pm 1.1 \pm 0.4 )
                      \times 10^{-4} \: {\rm fm}^3,$$
in agreement with a model-independent dispersion sum rule, and
$$\bar{\alpha}-\bar{\beta} = ( 10.8 \pm 1.1 \pm 1.4 \pm 1.0 )
                      \times 10^{-4} \: {\rm fm}^3,$$
where the errors shown are statistical, systematic, and model-dependent,
 respectively.
A comparison with previous experiments is given and global values for
the polarizabilities are extracted.
\end{abstract}

\narrowtext



\section{Introduction}
The electric and magnetic polarizabilities of the nucleon, labeled
$\bar{\alpha}$ and $\bar{\beta}$ respectively, are fundamental structure
constants that characterize the ability of the constituents
of the nucleon to rearrange themselves in response to static or slowly
varying external electric and magnetic fields.  These parameters are
as fundamental as the charge and magnetic radii of the nucleon, although
they have received considerably less attention until fairly recently.
With the high present-day interest in QCD-based descriptions of the structure
of the nucleon, the additional information represented by
an accurate determination of the polarizabilities would be of substantial
importance.  This question has motivated considerable activity in
recent years on both theoretical and experimental fronts.
In this Introduction, the experimental situation is reviewed;  a
discussion of the theoretical issues is presented in
Section~\ref{sec:discussion}.

Measurements of the proton polarizabilities have come exclusively
from Compton scattering experiments.  These measurements
rely on a theorem which establishes a unique relation
between the model-independent
low-energy expansion (LEX) of the Compton scattering cross section and
the polarizabilities.  For photon energies sufficiently low, this expansion
in the laboratory frame reads \cite{Pet:81}
\begin{eqnarray}
\lefteqn{
\frac{d\sigma}{d\Omega}(\omega,\theta)=\frac{d\sigma}{d\Omega}^{\rm B}
(\omega,\theta)  } \nonumber \\
& & -\frac{e^{2}}{4\pi M}\left(\frac{\omega'}
{\omega}\right)^{2}\left(\omega\omega'\right)
\left\{\frac{\bar{\alpha}+\bar{\beta}}
{2}(1+\cos \theta)
^{2}+\frac{\bar{\alpha}-\bar{\beta}}{2}(1-\cos \theta)^{2}\right\},
\label{eq:lex}
\end{eqnarray}
where $\omega$ and $\omega^\prime$ are the energies of the incident
and
scattered photons respectively, $e^2/4\pi$ is the fine structure constant,
and $d\sigma^{\rm B}/d\Omega$ is the exact
Born cross section
for a proton with an anomalous magnetic moment but no other
structure \cite{Pow:49}.  The
LEX is an expansion of the cross section to first order in
$(\omega\omega^\prime)$; besides the anomalous magnetic moment, the only
structure-dependent terms to this
order are
the polarizabilities.
The equation shows that the forward and backward cross sections
are sensitive mainly
to $\bar{\alpha}+ \bar{\beta}$ and
$\bar{\alpha}-\bar{\beta}$, respectively, whereas the 90$^\circ$
cross section is sensitive only to $\bar{\alpha}$.
The sum $\bar{\alpha}+ \bar{\beta}$ is independently constrained by a
model-independent dispersion sum rule \cite{Bal:60}:
\begin{equation}
\bar{\alpha}\,+\,\bar{\beta}=\frac{1}{2\pi
^{2}}\int_{m_{\pi}}^{\infty}
\frac{\sigma_{\gamma}(\omega)d\omega}{\omega^{2}}\,=\,14.2\, \pm\,
0.5,
\label{eq:sumrule}
\end{equation}
in units of $10^{-4}$ fm$^3$  (these units are implicitly
understood hereafter),
where $\sigma_{\gamma}(\omega)$ is the total photoabsorption cross
section on the proton.  The numerical value in Eq.~\ref{eq:sumrule}
is obtained using both
the available experimental data and a reasonable theoretical
{\it ansatz} for extrapolating the integral to infinite
energy \cite{Dam:70,Lvov:79}.

The polarizabilities are determined by measuring the deviation of the cross
sections from the Born values.
A plot of the Born and LEX cross sections is shown in
Fig.~\ref{fig:cross_section_plot}.  The
curves show that the effect of the polarizabilities, which is proportional to
the difference between
the Born and LEX curves, is
not very large,
thereby placing great demand on the
statistical precision and systematic accuracy of the measurements
in order to
obtain precise values for the polarizabilities.
Those demands can be relaxed by going to a higher photon energy, since
the sensitivity of the cross section
to the polarizabilities increases
with energy.
However if the energy becomes too large,
the LEX breaks down and
theoretical uncertainty is introduced into the extraction of the
polarizabilities from the measured cross sections.  This is demonstrated in
Fig.~\ref{fig:cross_section_plot}, where the curve labeled DR is a
 calculation of the cross section using dispersion relations and is valid,
in principle, at all energies.
This curve shows that the LEX is not valid above about 100 MeV.
Since many of the experiments, including the present
measurements, have been done
at energies
outside the range of validity of the LEX, it is necessary to pay particular
attention to the
model dependence in the determination of the polarizabilities from the cross
sections.  This
issue is discussed at length in Section~\ref{sec:theoretical}.

The above considerations suggest that the experimental challenge for
determining
the proton polarizabilities is to measure an {\it absolute} cross
section over an
energy and angular range that is appropriately balanced between
sensitivity to the polarizabilities and
insensitivity to any theoretical model.
With this in mind, it is useful to examine briefly the previous
experiments that have attempted to determine
the proton polarizabilities from
measurements of the Compton scattering cross
section \cite{Go60,Baranov:74,Federspiel:91,Zieger:92,Hallin:93},
some of which are summarized in Table~\ref{tab:experiments}.

The pioneering experiment is that of Gol'danski \cite{Go60}, whose
original result ($\bar{\alpha}=9\pm 2$) is omitted from the list mainly
because the 8\% uncertainty in the normalization of the cross sections leads to
a systematic uncertainty in $\bar{\alpha}$ of $\pm 5$ \cite{Bern74}.
Also omitted is an earlier Compton scattering
experiment \cite{Ox58} with  energies, angles, and
systematic errors comparable to those of Gol'danski,
although no attempt was made by the author
to extract polarizabilities from the cross sections.
In both cases the systematic uncertainty on the polarizabilities
is too large to affect the global fit we will report in
Section~\ref{sec:extraction}.

The Moscow 1975
experiment \cite{Baranov:74} used a bremsstrahlung photon beam and
a photon detector with very poor energy resolution.
Nevertheless, due to a clever technique that allowed the cross section to
be normalized to the well-known Klein-Nishina cross section for Compton
scattering on the electron, the systematic errors were thought to be small.
The reported polarizabilities were based on a fit to the cross
sections using the LEX.  Since the maximum energy of those cross sections
(110 MeV) is outside the range of validity of the LEX, we have refitted them
using the dispersion-relation
technique described in Section~\ref{sec:extraction},
and those results are given in Table~\ref{tab:experiments}.
Unfortunately they are badly
inconsistent with the dispersion sum rule, thereby casting
doubt on the cross sections and providing the principal
motivation for the more recent experiments.

The Illinois 1991 experiment \cite{Federspiel:91}
had two very desirable features.  First,
it was done at both a forward and backward
scattering angle and at low energies, so that
model-independent determinations of both
$\bar{\alpha}+\bar{\beta}$ (testing the dispersion
sum rule and/or the systematics
of the experiment) and
$\bar{\alpha}-\bar{\beta}$ were possible.  Second,
a tagged photon beam was used,
thereby considerably improving the ability to measure absolute
cross sections accurately.
Unfortunately, the combined effects of low energy (implying
low sensitivity) and
the counting rate limitations inherent in
a  tagged photon experiment resulted in reduced
statistical precision
in the extracted polarizabilities.

The Mainz 1992 experiment \cite{Zieger:92} measured the 180$^\circ$
Compton cross section by detecting the recoil
proton at 0$^\circ$ in a magnetic spectrometer,
normalizing to the Compton cross section on the electron.
The energy, 132 MeV,  was a good compromise between
sensitivity to the polarizabilities and model independence.  The principal
drawback of this experiment was that
$\bar{\alpha}-\bar{\beta}$ was determined by just
a single cross-section
measurement at one energy and angle.\footnote{Actually
cross sections were measured at both 132 and 98 MeV, but the
the latter datum has poor statistical quality and does not provide
a serious constraint on the polarizabilities.}

The Saskatoon 1993
experiment \cite{Hallin:93} used a high duty-factor bremsstrahlung photon
beam and
a high-resolution NaI detector to measure
an extensive set of angular distributions via the endpoint technique.
While the statistical and systematic quality of the data were very good,
the  energies (150--300 MeV) were far outside the range of validity of
the LEX, leading to possibly large
uncertainties in the polarizabilities due to model-dependent effects.

We report here new measurements of the Compton scattering cross sections
and the extraction of improved values for the
electric and magnetic polarizabilities from those cross sections.
The measurements utilized a new experimental technique, described in
Sections~\ref{sec:experiment} and \ref{sec:reduction},
in which measurements were done simultaneously using tagged photons
(70--100 MeV) and untagged photons (100--148 MeV).
An important feature of this work is a careful consideration of the model
dependence in the extraction of the polarizabilities from the cross sections;
the theoretical background for this discussion
is presented in Section~\ref{sec:theoretical}.
In Section~\ref{sec:extraction},
the actual procedure used to extract the polarizabilities is presented,
and the new values derived from both the present experiment and
a global fit to all the recent experiments are given.
A brief discussion of the
theoretical impact of the results is given in
Section~\ref{sec:discussion} and our conclusions are summarized in
Section~\ref{sec:conclusion}.  An extensive account of this work can be found
in the Ph.D. dissertation of MacGibbon \cite{MacG:94}.
\section{Experiment}
\label{sec:experiment}
The Compton scattering cross section for the proton was measured at the
Saskatchewan Accelerator Laboratory (SAL) using tagged photons with
energies from 70 to
100~MeV and untagged photons with energies from 100 to 148~MeV.
The entire energy region
70--148~MeV was measured simultaneously.
A high duty-factor ($\sim$70\%) beam of 148-MeV electrons was
incident on a 115-$\mu$m aluminum radiator, creating a
continuous
bremsstrahlung spectrum of photons up to the endpoint.
The electron beam
energy was chosen
to minimize the background from the $\pi^0$-decay
photons.
The photon beam was collimated so that its diameter was
$\sim$6~cm at the target position.
Approximately 25\% of the
photon beam was removed by the collimation.
The primary electron beam was bent away from the photon beam by
the main tagging magnet and steered into a shielded beam dump by a secondary
magnet.

The SAL tagger\cite{SALtagger:93} was used to determine the energy of the
tagged
photons by momentum analyzing the associated electrons in the tagging magnet
and detecting them in
a 62-channel scintillator hodoscope in coincidence with the scattered photons.
With the electron beam energy of 148~MeV, the magnetic field was
set to tag 70--100~MeV photons with an average resolution of
$\sim$0.5~MeV for each of the 62 channels.
The total tagged photon flux, integrated over the 62 channels, was
approximately
$7\times 10^{7}$ sec$^{-1}$.

The scattering target was liquid hydrogen.  The target assembly was
a closed-loop
recirculation system consisting of a two-stage refrigerator, a target flask and
reservoir, and transfer tubes.
The 0.25-mm thick Mylar
target flask was a cylinder 10.16~cm in diameter and $\sim$13~cm long,
oriented with the symmetry axis along the beam direction.
After the initial cooldown and liquification, the flask could be remotely
emptied or filled in $\sim$20 minutes, thereby
facilitating the process of alternating between full-target and empty-target
runs.  The flask was contained in an aluminum vacuum chamber;
the detectors were
shielded from the photons scattered from the
0.25~mm Mylar windows isolating the vacuum chamber from the atmosphere.

By measuring the transmission of photons
through the target, its thickness was determined {\it in situ}.
A tightly collimated beam of 0.661-MeV photons from a $^{137}$Cs source
was used.  By counting the number of photons transmitted
by both the full and empty target and by using the
known mass absorption
coefficient of hydrogen \cite{Hubbell:82},
the target thickness was
determined to be
$13.00 \pm 0.15$~cm, or $(5.43 \pm 0.06) \times 10^{23}\, {\rm
protons} /{\rm cm}^2$, in agreement with the geometrical measurement.

The photons were detected in two large-volume high-resolution NaI(Tl)
spectrometers, one each at scattering angles of 90$^\circ$ and 135$^\circ$
and each subtending a geometrical solid
angle of $\sim0.05$ sr.  The NaI crystals were surrounded by plastic
anti-coincidence shields to reject cosmic rays and other charged particles and
by 10.2~cm-thick Pb
shields to reduce the effects of room background.

The gains of the detectors were monitored during data acquisition
and corrected for drifts over the
course of the three-week run.  An LED
was mounted on the back of each NaI detector together with
an annular silicon surface-barrier detector (SBD), so that both the SBD
and the NaI were illuminated by photons from the LED.
The LED was flashed at two different intensities, allowing a monitor of
both the gain and offset of each photomultiplier tube (PMT)
 and its associated electronics.
The SBD, which was used to monitor and correct the intensity of the LED pulses,
was determined to be stable to better than 0.5\% in bench tests
with an $\alpha$-particle source.
The corrected gain of each NaI was stable to better than 0.4\% throughout the
course of the experiment.

The overall trigger for the experiment was a signal with an
energy above threshold
in either of the two NaI detectors.  All such events
were accepted whether or not they were accompanied by a tagging electron,
thereby allowing the acquisition of
both tagged and untagged events simultaneously.
For each event, the ADC value for each of the NaI PMT's was recorded
along with
the 62 TDC values corresponding to the time difference between a NaI signal and
each of the tagging electrons.  At 15-second intervals,
the 62 scalers counting the number of electrons in each tagging channel
were read out.
The data were sorted off-line into tagged and
untagged events.  A tagged event was defined
as one in which the photon was correlated with a tagging electron, and an
untagged event was defined as {\it any} photon event.  Thus, the tagged
events were a {\it subset} of the untagged events.

Data were collected by alternating between
full-target runs of $\sim6$ hours and
empty-target runs of $\sim4$ hours.
At three different times during the experiment,
each NaI detector was placed directly in the photon beam in order to measure
the
ratio of the number of photons incident on the target
to the number of tagging electrons
(the so-called tagging efficiency) and to determine
the detector response functions (see Section~\ref{sec:reduction}) and
energy calibration.
A Pb-glass detector was also periodically
placed in the photon beam in order to provide a
relative monitor of the tagging efficiency, which was determined to be stable
to
better than $\pm~0.5\%$ throughout the course of the experiment.


\section{Data Reduction}
\label{sec:reduction}
Since the present technique of simultaneous measurements with tagged and
untagged photons is new, it is discussed here in some detail.  The technique
is shown schematically in Fig.~\ref{fig:photon} and consists
of two types of measurements:  (1) a scattering run,
which measures the
detector pulse-height spectrum for scattered photons
normalized to the number of electrons $n_e$ detected
in the tagger; and (2) a calibration run which is used to normalize $n_e$
to the number of photons incident on the target.  The latter is accomplished
by putting the detector
directly into the photon beam.
Together these measurements are used to determine the scattering
cross section.

For both the tagged and untagged data, the scattering cross sections were
determined from the detector pulse-height spectrum of scattered photons:
\begin{equation}
   \frac{dN_{\rm s}(E)}{dE} \, = \,
\int_0^{E_0} \frac{dN_{\rm i}(E_\gamma)}{dE_\gamma} \,
              \kappa \Omega \frac{d\sigma(E_\gamma)}{d\Omega} \,
             \frac{dR_{\rm s}(E_\gamma,E)}{dE} \,dE_\gamma,
\label{eq:int_scatt}
\end{equation}
where
$\kappa$ is the target thickness, $\Omega$
is the detector solid angle,
$E_\gamma$ is the incident
photon energy, $E$ is the pulse height in the detector, and
$E_0$ is the endpoint energy.
This expression is the convolution of the bremsstrahlung
photon spectrum incident on the target
$dN_{\rm i}(E_\gamma)/dE_\gamma$, the scattering cross section
$d\sigma(E_\gamma)/d\Omega$, and the detector response function for
scattered photons $dR_{\rm s}(E_\gamma,E)/dE$.
This latter quantity is
the probability density that a photon of energy $E_\gamma$ incident on
the front
face of the NaI detector will register a pulse height $E$ in that
detector.\footnote{Strictly speaking,
the first argument of the scattering
response function should be $E_\gamma^\prime$,
the scattered photon energy, which differs from the incident energy due to
the Compton recoil shift.  For simplicity of notation, this recoil shift
has been absorbed into the response function.}
The detector pulse-height spectrum
in a calibration run is used to measure the incident
photon flux:
\begin{equation}
   \frac{dN_{\rm c}(E)}{dE} \, = \,
 \int_0^{E_0} \frac{dN_{\rm i}(E_\gamma)}{dE_\gamma} \,
             \frac{dR_{\rm c}(E_\gamma,E)}{dE} \,dE_\gamma,
\label{eq:int_calib}
\end{equation}
where the
detector response function in the calibration run
$dR_{\rm c}(E_\gamma,E)/dE$ differs slightly from that in the scattering
run $dR_{\rm s}(E_\gamma,E)/dE$ due to the different
geometries and the Compton recoil shift.

\subsection{Tagged Photon Analysis}
For the tagged-photon measurements, each incident tagged-photon energy bin
is a 0.5-MeV slice of the full bremsstrahlung spectrum,
or essentially a $\delta$-function:
\begin{equation}
\frac{dN_{\rm i}(E_\gamma)}{dE_\gamma} \, \approx \, n_{\rm t}(E_t)\,
\delta(E_\gamma-E_t),
\label{eq:delta}
\end{equation}
where $E_t$ is the tagged-photon energy and
$n_{\rm t}$ is the number of tagged photons in that bin.
  Therefore the pulse-height spectrum for
a calibration run is
given by
\begin{equation}
   \frac{dN_{\rm c}(E)}{dE}
       =  n_{\rm ec}\, \epsilon_t \,
	\frac{dR_{\rm c}(E_t,E)}{dE}, \label{eq:calib}
\end{equation}
where
$n_{\rm ec}$ is the number of associated tagging electrons measured with
the tagger scalers, and
$\epsilon_t \equiv n_{\rm t}(E_t)/n_{\rm ec}$ is the tagging efficiency.
This latter quantity is determined experimentally
by taking the ratio of the total number
of counts in the pulse-height spectrum to $n_{ec}$.
Note that the tagged-photon pulse-height spectrum is a direct measure of
the shape of
the detector
response function in the calibration mode.
A typical spectrum is shown in Fig.~\ref{fig:tagged_brems} which also
shows a Monte Carlo simulation of the response function using the code
EGS4\cite{egs4}.
The excellent agreement between the calculated and measured response functions
for the calibration geometry
gives us confidence that EGS4 can be used to calculate the detector response
for the scattering geometry, where it is not so easily measured.
Previous experience with the same detectors at similar photon energies has
shown that EGS4 {\it can} accurately account for scattering line shapes in
experiments where they can be measured with good statistics
\cite{Mellendorf:93}.

Again using Eq.~\ref{eq:delta},
the detector pulse-height spectrum for tagged photons in a scattering run is
related to the cross section by
\begin{equation}
  \frac{dN_{\rm s}(E)}{dE}  = n_{\rm es} \, \epsilon_t \, \kappa \Omega \,
                  \frac{d\sigma}{d\Omega}(E_t)\,\frac{dR_{\rm s}(E_t,E)}{dE},
\end{equation}
where $n_{\rm es}$ is the number of tagging electrons measured with the tagger
scalers, and the value of $\epsilon_t$ is obtained from the calibration run.
The detector response function is again determined using an EGS4 simulation,
which now also
accounts for the effects of the finite geometry
on $\kappa$ and $\Omega$, as well
as the Compton recoil shift.  For each tagger
channel, cuts are placed on the
TDC value (see Fig.~\ref{fig:tdc}) to obtain separate
true-plus-random and random pulse-height spectra.  The appropriately
normalized random spectrum, which is corrected for
rate-dependent losses in the TDC,
is then subtracted from the true-plus-random one to
obtain a
true coincidence spectrum, which is then
shifted to a common energy and combined with those of adjacent
tagger channels into one of four composite spectra,
each
corresponding to a tagging range of about 8 MeV.  This is done separately for
the full-target and empty-target data, which are then subtracted to obtain
final spectra for scattering from hydrogen.\footnote{A 2\% correction
is made for the residual gas in the target flask during
the empty-target runs.}
These spectra
are then integrated to obtain
$d\sigma/d\Omega$.  A
sample hydrogen scattering spectrum together with a normalized EGS4 spectrum
is shown in Fig.~\ref{fig:tagged_H}.
The resulting tagged-photon
cross sections are given in the Appendix.

\subsection{Untagged Photon Analysis}
The untagged cross sections were also determined using
Eqs.~\ref{eq:int_scatt} and \ref{eq:int_calib}.
To simplify
the notation, the incident photon spectrum is expressed as
\begin{eqnarray}
  \frac{dN_{\rm i}(E_\gamma)}{dE_\gamma}
  & = & n_{\rm 0} \, f(E_\gamma) \nonumber \\
  & = &  n_{\rm ec} \, \epsilon_u \, f(E_\gamma),
\end{eqnarray}
where $n_{\rm 0}$ is the total
number of incident photons with energy
in the interval between
$E_1$=100 MeV and the endpoint energy $E_0$,
$f(E_\gamma)$ is the spectrum shape that is
normalized to unit area over the same energy interval, and
$\epsilon_u$ is
the ratio of $n_{\rm 0}$ to the number of electrons measured by the
tagger scalers.  Thus for the untagged analysis, the tagger is only used to
normalize the number of photons in a calibration run to that in a scattering
run.  The pulse-height spectrum in a calibration run
(see Eq.~\ref{eq:int_calib})
now takes the form
\begin{equation}
   \frac{dN_{\rm c}(E)}{dE}
    =  n_{\rm ec} \epsilon_u \int_{E_1}^{E_0} f(E_\gamma) \,
             \frac{dR_{\rm c}(E_\gamma,E)}{dE} \,dE_\gamma.
\end{equation}
The quantity $\epsilon_u$ is determined by normalizing the calibration
pulse-height spectrum to a calculated spectrum.  In the calculation,
the detector response function is again determined using EGS4,\footnote
{For the untagged analysis, it was necessary to have detector
response functions up to 148 MeV.
Since the maximum tagged photon energy in the setup was 100 MeV,
a supplemental experiment was performed {\it in situ} with higher-energy
 electrons
in order to measure the detector response.  It was verified that
the EGS4 simulation continues to describe the measured response function
up to an energy of 148 MeV.}
and the shape $f(E_\gamma)$ is taken to be the
Schiff bremsstrahlung spectrum \cite{Schiff:51}, which is
differential in both energy and angle.  The
resulting normalized spectrum for one of the detectors
is
shown in
Fig.~\ref{fig:untagged_brems} along with the data.
It was verified that the cross
sections derived from this technique are insensitive to
the exact form of $f(E_\gamma)$.

The cross sections are found by fitting the scattering pulse-height
spectrum to the expression (see Eq.~\ref{eq:int_scatt}):
\begin{equation}
 \frac{dN_{\rm s}(E)}{dE} \, = \,
 n_{\rm es} \epsilon_u \int_{E_1}^{E_0} f(E_\gamma) \,
             \kappa \Omega \frac{d\sigma}{d\Omega}(E_\gamma) \,
             \frac{dR_{\rm s}(E_\gamma,E)}{dE} \,dE_\gamma,
\end{equation}
using a simple parametrization of the cross section:
\begin{equation}
  \frac{d \sigma}{d \Omega}(E_\gamma)  =
\left\{
\begin{array}{ll}
             {\rm S}_1 &\mbox{  for $E_\gamma$ = 100--110  MeV} \\
             {\rm S}_2 &\mbox{  for $E_\gamma$ = 110--120  MeV} \\
             {\rm S}_3 &\mbox{  for $E_\gamma$ = 120--130  MeV} \\
             {\rm S}_4 &\mbox{  for $E_\gamma$ = 130--140  MeV} \\
             {\rm S}_5 &\mbox{  for $E_\gamma$ = 140--150  MeV}.
\end{array}
\right.
\end{equation}
The  S$_{\rm i}$ were allowed to vary freely to obtain the best fit
to the final scattering spectrum (with empty-target subtracted) over
the energy range
corresponding to incident photons between 100 MeV and the endpoint energy.
These parameters have a high degree of anti-correlation between adjacent
values due to the low-energy tail on the detector response function; therefore
it was necessary to employ the full error matrix when using these
cross sections to extract physical quantities.
The scattering spectra and the associated
fits are shown in Fig.~\ref{fig:untagged_H}.  The resulting untagged cross
sections and associated errors are given in the Appendix.

It is noted that since the tagged data are a subset of the
untagged data, the minimum incident
energy for which untagged results are quoted
(100~MeV) is the maximum energy for which we have tagged results.  Thus
the same photon event is not counted as both tagged and untagged.
However,
as a check of our technique, we
investigated the overall consistency between the tagged and untagged
cross sections by extending the calculation shown in Fig.~\ref{fig:untagged_H}
into the tagging region using the tagged cross sections given in the Appendix.
At 135$^\circ$ there is good consistency.  However, at 90$^\circ$ the
curve is consistently below the data, indicating that those data
are contaminated with
non-Compton events from the target, such as $\pi^0$-decay photons, whose
kinematic endpoint is just below the maximum tagging energy of 100 MeV, or
pair production followed by bremsstrahlung.  This is not a
problem for energies above the tagging region.

\subsection{Systematic Errors}

The overall systematic uncertainty for the cross sections is $\pm$2.9\% for
the untagged data and ranges from $\pm 3.0$\% to $\pm 4.0$\% for the
tagged data.  Typical contributions include the uncertainties in
the detector solid angle
($\pm$1.2\%), target thickness ($\pm$1.2\%), tagging efficiency
($\pm$1--2\%), and rate-dependent corrections ($\pm$1--3\%).
A complete discussion
of these errors and the details of the experimental procedure and
data reduction have been given by MacGibbon \cite{MacG:94}.
It should be noted that the different systematic errors had various degrees of
correlation among the different cross sections (e.g., the error due to
the target thickness was correlated among all the cross sections, while that
due to the solid angle of the 135$^\circ$ detector was correlated only
among the 135$^\circ$ cross sections).  The effect of these correlations was
properly taken into account when extracting the polarizabilities.


\section{Theoretical Considerations}
\label{sec:theoretical}

Since much of the data in the current experiment lie
outside the range of validity of the
LEX, it is necessary to address the issue of how to extract the
polarizabilities from
the cross sections and the model dependence therein.
In this section, a theoretical overview of the dispersion-relation
approach utilized here is presented.
This approach, as well as the computer code used for
the numerical work, are due to L'vov \cite{L'vov:81}.
A more complete
description will be published elsewhere \cite{L'vov:95}.

All Compton scattering observables are determined from
six independent invariant
amplitudes, $A_{i}(\nu,t)$, which are free of kinematical singularities and
constraints and are
even functions of $\nu=(s-u)/4M$.
Here $s$, $t$, and $u$ are the usual Mandelstam variables and $M$ is
the proton mass.
These amplitudes incorporate different mechanisms for Compton scattering,
including $s$-,
$u$-, and  $t$-channel exchanges of stable particles (see diagrams in
Fig.~\ref{fig:diagrams}).
The $s$- and $u$-channel processes
are sums over all possible intermediate states that can
be formed in the photon-proton interaction, such as nucleon resonances,
non-resonant $\pi$-nucleon states, etc.  For the special case where the
intermediate state is a proton in the ground state,
the so-called
Born amplitude, $A_i^{\rm B}$, is obtained.  This purely real amplitude
is calculated exactly in terms of the charge,
mass,
and magnetic moment of the proton; the cross section based on
the $A_i^{\rm B}$ alone
is shown in Fig.~\ref{fig:cross_section_plot}.
All other diagrams, including
the $t$-channel exchanges, contribute to the non-Born part,
$A_i^{\rm NB}$.  At low energies, the Born and
non-Born parts combine to give the LEX,
with the polarizabilities identified as:
\begin{eqnarray}
\bar{\alpha} \, + \,\bar{\beta} \,
  & = & -\left( \frac{1}{2\pi} \right)
  \left[ A_{6}^{\rm NB}(0,0) + A_{3}^{\rm NB}(0,0) \right] \nonumber \\
\bar{\alpha} \, - \, \bar{\beta} \, & = & \, -\left(\frac{1}{2\pi}\right)
                       A_{1}^{\rm NB}(0,0).
\label{eq:ab_disp}
\end{eqnarray}

If $A_i$ falls sufficiently rapidly with increasing energy, it
satisfies an unsubtracted
fixed-$t$ dispersion relation:
\begin{equation}
 {\rm Re} \left[ A_i(\nu,t) \right]  \,=\,  A_i^{\rm B}(\nu,t) \,+\,
 \frac{2}{\pi}{\cal P}
   \int_{\nu_0}^{\infty} \frac{ \nu^\prime \, {\rm Im} \left[
A_i(\nu^\prime,t) \right]}
    {\nu^{\prime 2} - \nu^2} d\nu^\prime,
\label{eq:gen_disp}
\end{equation}
where $\nu_0$ is the threshold for pion photoproduction.
The imaginary parts of the scattering
amplitudes are related to the multipole amplitudes for the
total photoabsorption cross section on the proton through unitarity.
Therefore, provided the dispersion relations are valid,
a complete knowledge of the multipole amplitudes at all energies
uniquely determines the scattering cross section and therefore the
polarizabilities.
However for some of the amplitudes an unsubtracted dispersion relation is
{\it not valid} because the integral does not converge.
This is largely due to the $t$-channel
processes (see Fig.~\ref{fig:diagrams}), which give rise to amplitudes
that do not fall rapidly enough with energy to assure convergence of
the integral.  Indeed, the $t$-channel exchange of a stable particle
such as a $\pi^0$
leads to an amplitude which is independent of $\nu$
and unconstrained by
the multipole amplitudes (see Eq.~\ref{eq:tchan}).
For such amplitudes, Eq.~\ref{eq:gen_disp}
is not valid and consequently the scattering cross section is not
uniquely determined by the photoabsorption cross section.

Different approaches have been used to handle the convergence problem
\cite {L'vov:81,Filkov:81,Guiasu:78b}.
The approach utilized by L'vov \cite{L'vov:81} is to
terminate the integral at $\nu_m$ = 1.5 GeV and close
the contour with a semicircle $\cal C$ of radius $\nu_m$
in the upper half of the complex $\nu$-plane.  The dispersion relation then
takes the form
\begin{equation}
 {\rm Re} \left[ A_i(\nu,t) \right] =
A_i^{\rm B}(\nu,t) + A_i^{\rm int}(\nu,t) + A_i^{\rm asymp}(\nu,t),
\label{eq:disp_int_asymp}
\end{equation}
where the integral part $A_i^{\rm int}$ is given by
\begin{equation}
A_i^{\rm int}(\nu,t) = \frac{2}{\pi}{\cal P}
   \int_{\nu_0}^{\nu_m} \frac{ \nu^\prime \, {\rm Im} \left[ A_i(\nu^\prime,t)
\right]}                    {\nu^{\prime 2} - \nu^2}d\nu^\prime,
\label{eq:int}
\end{equation}
and the asymptotic part $A_i^{\rm asymp}$ is given by
\begin{equation}
   A_i^{\rm asymp}(\nu,t) = {\rm Im}\left[\frac{1}{\pi} \int_{\cal C}
                 \frac{\nu^\prime \, A_i(\nu^\prime,t)}
                               {\nu^{\prime 2} - \nu^2}d\nu^\prime \right].
\label{eq:asymp}
\end{equation}
It should be emphasized that this procedure is exact and does not rely
on any special assumptions about the behavior of the amplitudes at
very large energies.  It expresses the real part of the invariant amplitude
as a sum of a Born part which is calculated exactly, an integral part which
is determined by the photoabsorption multipole amplitudes, and an
asymptotic part.
The integral and asymptotic
contributions are now discussed separately.

The integral part is discussed first.
As mentioned above,
unitarity relates Im$[A_i]$ to the multipole amplitudes for
the total photoabsorption cross section on the proton.
For Compton scattering near and below the pion threshold, the most
important intermediate states contributing to the dispersion integrals are the
$\pi N$ states (including nucleon resonances that decay into the $\pi N$
channel), which have both the largest cross section and the largest
energy weighting.
For this contribution, Im$[A_i]$ is a sum of
bilinear
combinations of single-pion photoproduction multipole amplitudes, the most
important of which have been very well measured and
tabulated \cite{MW:75,VPI:90}.  Thus the $\pi N$
contribution to the dispersion integrals can
be reliably calculated, and the sensitivity
of the scattering cross section to experimental
uncertainties in the $\pi N$ multipoles can
be readily determined.
On the other
hand, the multipole amplitudes for multi-pion photoproduction, which
dominates the photoabsorption cross section for energies greater than
$\sim$ 600 MeV, are poorly
known experimentally and therefore can only be treated in the context of a
model.  Fortunately, for photon scattering energies below 150 MeV, the
multi-pion contribution to the dispersion integrals is suppressed
because of the large energy denominator,
so it is expected that the predicted scattering cross sections will
only weakly depend on the model assumptions.  Approximately 40\% of
the multi-pion cross section proceeds either
through known resonances or through the $\rho N$ channel, and
these contributions
can be reliably decomposed into multipoles.
The non-resonant cross section is partly due to $\pi\Delta$ production.  For
partial waves $l\ge 1$, the multipole amplitudes for this process are
calculated in the Born approximation, assuming a one-pion-exchange
mechanism.  The remaining multi-pion cross section, hereafter
referred to as the non-resonant s-wave contribution, is ascribed some
combination of low-order multipoles,
leading to non-resonant intermediate states with
$j^\pi=1/2^+$, $1/2^-$, or $3/2^-$,
and then scaled so that the calculated total multi-pion
contribution to the photoabsorption reproduces the
experimental value at each energy.
The sensitivity of the scattering cross
section to the model can be tested by
adjusting the combination of multipoles making up
the non-resonant s-wave contribution.

The asymptotic part is discussed next.
Using Regge theory, L'vov
shows \cite{L'vov:81} that for $i=$3--5,
$A_i(\nu,t)$ drops sufficiently rapidly at high $\nu$
and fixed
$t$ to assure negligible contribution of $A_i^{\rm asymp}$
for $\nu_m$ = 1.5 GeV.
However, both
$A_1$ and $A_2$ and to a lesser extent $A_6$
have important contributions due to $t$-channel
exchanges, so that the
asymptotic part cannot be neglected.
For $\nu\ll\nu_m$, the asymptotic amplitude
due to the $t$-channel
exchange of a particle $x$ is $\nu$-independent and has
the general form (see Fig.~\ref{fig:diagrams})
\begin{equation}
A_i^{\rm asymp}(t)\,=\,-\frac{g_{xNN}f_{x\gamma\gamma}}{t-m_x^2}\,F(t-m_x^2),
\label{eq:tchan}
\end{equation}
where $g_{xNN}$ and $f_{x\gamma\gamma}$ are
the $N\rightarrow Nx$ and $x\rightarrow
\gamma\gamma$ coupling constants respectively, $F$
is the product of vertex form factors,
and $m_x$ is the mass of the exchanged particle.
This leads to a $\nu$-independent
amplitude whose $t$-dependence is mainly determined by $m_x^2$.
In reality there may be
a whole family of exchanged particles, leading
to a more complicated $t$-dependence.
Nevertheless, for the low $t$ of the
scattering data considered here ($|t|\le 0.06$ GeV$^2$),
the amplitude is completely dominated by the
lowest-mass exchanged particle, and the
effect of higher-mass exchanges can be absorbed into the form factor $F$.

For $A_2$ the lowest-mass exchanged
particle is a $\pi^0$, for which the coupling constants are
known experimentally to within approximately $\pm 4\%$ \cite{PDG,Nijmegan}.
The vertex form
factor is of the form $F=e^{B_\pi(t-m_\pi^2)}$, with $B_\pi$ in the range
0--3 GeV$^{-2}$ \cite{L'vov:95,Nijmegan}.
Therefore, up to uncertainties in the coupling constants and the form factor,
$A_2^{\rm asymp}$ can be calculated, and the effect of the uncertainties on
the scattering cross
sections can be tested.

For $A_1$ the leading exchange is that
due to a correlated pair of pions in a 0$^+$
isoscalar state,
for which the mass and the couplings are poorly known.
For low $t$, it is convenient to express $A_1^{\rm asymp}$ in
the alternate form
\begin{equation}
A_1^{\rm asymp}(t)\,=\,C_{2\pi}e^{B_{2\pi}t/2},
\label{eq:disp_asymp_1}
\end{equation}
where $B_{2\pi}\approx 2/m_{2\pi}^2$
and $C_{2\pi}=-g_{(2\pi)NN}f_{(2\pi)\gamma\gamma}/m_{2\pi}^2$.
The
parameter $B_{2\pi}$ can be estimated either from the systematics of the
$t$-dependence of Compton scattering in the vicinity of
E = 1.5 GeV, from which $B_{2\pi} \approx$ 6 GeV$^{-2}$ \cite{Duda:83},
or from the Regge
parametrization of $pp$-scattering polarization data,
from which $B_{2\pi} \approx$
10 GeV$^{-2}$ \cite{L'vov:81,Berger:78}.  This
range for $B_{2\pi}$ (6--10 GeV$^{-2}$)
corresponds to $m_{2\pi}$ in the range 447--577 MeV,
in agreement with expectations based
on phenomenological descriptions of the $NN$
interaction \cite{VIJAY}.  It is also consistent with
the $t$-dependence calculated via a backward dispersion relation,
using the physical amplitudes
for the process
$\gamma\gamma\rightarrow\pi\pi\rightarrow N\bar{N}$ \cite{L'vov:95}.
Once again, the sensitivity of
the scattering cross sections to the value of $B_{2\pi}$ can be tested.
The remaining constant $C_{2\pi}$ is related  to the difference of the
polarizabilities, as
can be seen by combining and rearranging
 Eqs.~\ref{eq:ab_disp}, \ref{eq:disp_int_asymp}, \ref{eq:int},
and \ref{eq:disp_asymp_1}:
\begin{equation}
\bar{\alpha}\,-\,\bar{\beta}\,=\,-\left(\frac{1}{\pi^2}\right)
         \left\{\int_{\nu_0}^{\nu_m}{\rm Im}\left[A_{1}(\nu,0)\right]
         \frac{d\nu}{\nu} \, + \, \frac{\pi C_{2\pi}}{2}\right\}.
\label{eq:sumruled}
\end{equation}
In the analysis described below, $C_{2\pi}$ is treated as a free
parameter which is adjusted to fit the scattering cross sections.
This is equivalent
to treating $\bar{\alpha}-\bar{\beta}$ as a free parameter.

For $A_6$, it follows from Regge theory that the unsubtracted
dispersion relation,
Eq.~\ref{eq:gen_disp}, actually converges \cite{L'vov:81}, although it is
also known experimentally that
the integral is not saturated by 1.5 GeV \cite{Armstrong}.
For $\nu\ll\nu_m$, $A_6^{\rm asymp}$ is approximately
$\nu$-independent and has a $t$-dependence of the form \cite{L'vov:81,Rollnik}
\begin{equation}
A_6^{\rm asymp}(t)\,=\,C_{P}e^{B_{P}t/2},
\label{eq:disp_asymp_6}
\end{equation}
with $B_P$ in the range 6--10 GeV$^{-2}$ \cite{Duda:83,Rollnik}.
The remaining constant $C_P$ is related to
the sum of polarizabilities by
\begin{equation}
\bar{\alpha}\,+\,\bar{\beta}\,=\,\frac{1}{2\pi^2}
\int_{\omega_0}^{\omega_m}\frac{\sigma_\gamma(\omega)}{\omega^2}d\omega
\, - \, \frac{C_P}{2\pi}.
\label{eq:sumrules}
\end{equation}
By comparing the above equation with the sum rule (see Eq.~\ref{eq:sumrule}),
it is seen that
the term involving $C_P$ represents that part of
the sum rule that is missing when the
integral is terminated at $\omega_m$=1.5 GeV; numerically, it has the value
$C_P/2\pi \approx -0.9$.  The possibility that the
experimental value of $\bar{\alpha}+\bar{\beta}$ is
different from the sum-rule value can be investigated by
treating $C_P$ as a free parameter that is adjusted to fit the scattering data.

In summary, fixed-$t$ dispersion relations can be used to predict the
Compton scattering cross section below about 150 MeV in terms of two
parameters, $C_{2\pi}$ and $C_P$ (or equivalently $\bar{\alpha}-\bar{\beta}$
and
$\bar{\alpha}+\bar{\beta}$),
which are then adjusted
to fit the scattering data.
The calculations rely on the experimentally known
single-pion photoproduction multipole amplitudes \cite{MW:75,VPI:90},
a model for the multi-pion
photoproduction multipole amplitudes,
and certain assumptions about the asymptotic behavior.  Specifically,
it is assumed that $A_{3-5}^{\rm asymp}$ are negligible, that
$A_2^{\rm asymp}$ is dominated by $\pi^0$ exchange, and that
the {\it ansatzen} of
Eqs.~\ref{eq:disp_asymp_1} and \ref{eq:disp_asymp_6} are
valid for $A_{1}^{\rm asymp}$ and $A_{6}^{\rm asymp}$, respectively.
The principal sources of uncertainty are as follows:
\begin{itemize}
\item uncertainties in the single-pion multipole amplitudes
for $A_i^{\rm int}$
\item model uncertainties in the calculation of the multi-pion multipole
amplitudes for $A_i^{\rm int}$
\item experimental uncertainties in the $\pi^0$-exchange couplings for
$A_2^{\rm asymp}$
\item theoretical uncertainty in $B_\pi$, which modulates the $t$-dependence of
$A_2^{\rm asymp}$
\item theoretical uncertainty in $B_{2\pi}$, which governs the $t$-dependence
of
$A_1^{\rm asymp}$
\item theoretical uncertainty in $B_P$, which governs
the $t$-dependence of $A_6^{\rm asymp}$
\end{itemize}
The sensitivity of
the value extracted for
$\bar{\alpha}-\bar{\beta}$ to these aspects of the dispersion
calculation will be discussed in the next section.
The numerical work was done using variations of
the computer code {\sc gngn}, which was written and supplied by L'vov.

\section{Determination of the Polarizabilities}
\label{sec:extraction}
\subsection{Fitting Procedure}
\label{sec:procedure}
The polarizabilities were determined by fitting
the theoretical curves to the experimental cross sections, taking full account
of the statistical and systematic errors.
As remarked above, the statistical errors for the five individual untagged
cross sections for each angle are correlated, thereby
necessitating the use of the full $5\times 5$ error matrix.
The inclusion of the systematic errors in the fit is important since a 1\%
change in the overall normalization of every cross section results in a change
in the extracted value of $\bar{\alpha} - \bar{\beta}$ by approximately $0.5$.
To determine the effect of the systematic errors on the
polarizabilities, it is assumed that they are mainly errors of normalization
in the
measured cross sections.  A standard technique is used to account for
the systematic errors from different independent data sets \cite{Dagostini:94}.
For each data
set, $\chi^2$ is defined as follows:
\begin{equation}
\chi^2 \, = \, \left(\frac{\sigma^{\rm th}-
N\sigma^{\rm exp}}{N\epsilon}\right)^2 \, + \,
            \left(\frac{N-1}{\epsilon_N}\right)^2,
\label{eq:chisq}
\end{equation}
where $\sigma^{\rm exp}$ and $\epsilon$ are the experimental scattering
cross section and statistical error, respectively, and $\sigma^{\rm th}$ is the
corresponding calculated scattering cross section.
The second term of
Eq.~\ref{eq:chisq} takes into account the contribution of the
normalization to $\chi^2$, where
$N$ is the
normalization constant and
$\epsilon_N$ is the systematic error.
The total value of $\chi^2$ is obtained by
summing the $\chi^2$ for each data set.
Fits could be subjected to the sum-rule constraint
by including an additional data set consisting of a single
datum, whose experimental value and uncertainty are $14.2\pm 0.5$
(see Eq.~\ref{eq:sumrule}) and whose corresponding
calculated value is equal to $\bar{\alpha}+\bar{\beta}$.
Standard least-squares fitting
procedures were used to
adjust $\bar\alpha$, $\bar\beta$, and the normalization constants in order
to minimize the total $\chi^2$.
The net result is
that each data set is properly weighted based on its systematic
error, taking full account of the correlations in those systematic errors,
and the uncertainties in the fitted parameters
(i.e., the polarizabilities)
include contributions from both the statistical and the systematic errors.
The purely statistical contribution to those uncertainties
can be determined by
fitting with fixed values for the normalization constants.  The net
systematic contribution is derived by assuming the total uncertainty
is the result of
combining in quadrature the statistical and systematic contributions.
A detailed discussion of this technique is given by MacGibbon \cite{MacG:94}.

\subsection{Present Experiment}
\label{sec:present}

Fits were done using cross sections calculated with
the fixed-$t$ dispersion relations,
treating $\bar{\alpha}+\bar{\beta}$ and
$\bar{\alpha}-\bar{\beta}$ as free parameters.  In addition,
the sensitivity of the results derived for $\bar{\alpha}$ and $\bar{\beta}$
 to the uncertainties in the calculation was
investigated.  The results of this investigation are
presented in Table~\ref{tab:modeldep} and
a typical fit to the scattering cross sections is shown in Fig.~\ref{fig:fit}.
The sensitivity to the $\pi N$ multipole amplitudes
was studied by doing fits with
each of
four available tabulations of those amplitudes.
One was the Metcalf-Walker 1975
tabulation \cite{MW:75}; the remaining three were from the more recent VPI
tabulations \cite{VPI:90}, specifically the SP92, FA93, and SP95 data sets.
The sensitivity to the model assumptions in
the calculation of the multi-pion multipole
amplitudes was studied by adjusting the multipole composition of
the non-resonant s-wave contribution.
The sensitivity to the product of
coupling constants for the $\pi^0$ exchange
amplitude ($g_{\pi NN}f_{\pi\gamma\gamma}$)
 was studied by varying the product
  within the range of its experimental uncertainty, $\pm 4\%$.
 Finally, the sensitivity to each of the three parameters
describing the $t$-dependence of the asymptotic amplitudes
was studied by  varying them
within the ranges $B_\pi$=0--3, B$_{2\pi}$=6--10,
and $B_P$=6--10 GeV$^{-2}$.

Each of the model parameters in Table~\ref{tab:modeldep} was then chosen
to give results for the polarizabilities midway between the extreme
values.  The cross sections were fitted using these parameters in order
to obtain final results.  The total model uncertainty was taken to be
the combination-in-quadrature of the numbers in Table~\ref{tab:modeldep}.
The results thus obtained are
\begin{equation}
  \fbox{$\bar{\alpha} - \bar{\beta} =
               10.8 \pm 1.1 \pm 1.4 \pm 1.0$}
\label{eq:diff}
\end{equation}
and
\begin{equation}
  \fbox{$\bar{\alpha} + \bar{\beta} =
               15.0 \pm 2.9 \pm 1.1 \pm 0.4$}
\label{eq:sum}
\end{equation}
where the errors are statistical, systematic, and model-dependent,
respectively.
As noted in Table~\ref{tab:modeldep}, if the individual contributions to the
model-dependent errors are combined linearly
rather than in quadrature, then the model-dependent errors on
$\bar{\alpha} - \bar{\beta}$ and $\bar{\alpha}
+ \bar{\beta}$ become 2.0 and 0.7,
respectively.
Eqs.~\ref{eq:diff}-\ref{eq:sum} represent the final results of the present
experiment.

The result obtained for $\bar{\alpha} + \bar{\beta}$ agrees
with the sum-rule value of
$\bar{\alpha} + \bar{\beta} = 14.2 \pm 0.5$.  Moreover, it was verified that
the
result obtained for $\bar{\alpha} - \bar{\beta}$ is
independent of whether or not
the sum-rule constraint is applied in the fit.
The combination of
Eq.~\ref{eq:diff} and
the sum rule determines values for the individual polarizabilities:
\begin{equation}
   \bar{\alpha} =  12.5 \pm 0.6 \pm 0.7 \pm 0.5
\end{equation}
\begin{equation}
   \bar{\beta} = 1.7 \mp 0.6 \mp 0.7 \mp 0.5,
\end{equation}
where the individual errors
are anticorrelated due to the precise value of the sum.

\subsection{Global Average}
\label{sec:global}

The present results were then compared with those of the previous
experiments listed in Table~\ref{tab:experiments},
and ``global-average'' values for $\bar{\alpha}$
and $\bar{\beta}$ were determined.
As a first step, polarizabilities were extracted from each data set
separately and compared to
each other in order to check for overall consistency.
The dispersion cross sections were used to fit to each of
the data sets independently, without imposing the
sum-rule constraint.  The systematic errors were taken into account in the
manner described above, with the only other free parameters being
$\bar{\alpha}+\bar{\beta}$ and $\bar{\alpha}-\bar{\beta}$.
Model parameters were chosen to reproduce approximately the average.
In order to minimize the model dependence,
the Saskatoon 1993 cross sections \cite{Hallin:93}, all
of which are above the pion threshold,
were excluded from this analysis.
The results are given in Table~\ref{tab:experiments} and
are shown in Fig.~\ref{fig:ellipse}
in the form of error contours in
the $\bar{\alpha}\bar{\beta}$-plane, along with the constraint imposed by
the dispersion sum rule, Eq.~\ref{eq:sumrule}.
It is evident that there is excellent overall consistency among all but
the Moscow 1975 experiment \cite{Baranov:74}.  As already mentioned, it was
the inconsistency between that experiment and the sum rule that motivated the
more recent experiments.
The inconsistency can
be traced to their 150$^\circ$ cross sections, which lead to a value
of $\bar{\alpha}-\bar{\beta}$ that strongly disagrees with the other
experiments.  On the other hand, their
90$^\circ$ cross sections essentially determine $\bar{\alpha}$,
which appears to be consistent with the other experiments.
Nevertheless, in the determination of the global average,
none of the Moscow 1975 data were
considered.\footnote{We note that including the 90$^\circ$ Moscow data in the
global fit has essentially no effect on either the
inferred value of $\bar{\alpha}-\bar{\beta}$ or its uncertainty.}

The dispersion cross sections were used to fit
the remaining three data sets
without the sum-rule constraint, taking into account the
systematic errors and estimating the model uncertainty as described
earlier.  Global-average values for the polarizabilities are thus found:
\begin{equation}
\left(\bar{\alpha}-\bar{\beta}\right)_{\rm global} =  10.0 \pm 1.5 \pm 0.9
\label{eq:diff_global}
\end{equation}
and
\begin{equation}
\left(\bar{\alpha}+\bar{\beta}\right)_{\rm global} =  15.2 \pm 2.6 \pm 0.2,
\label{eq:sum_global}
\end{equation}
where the first error is the combined statistical and systematic error
propagated from the
individual cross sections and the second
is the model-dependent error derived in the manner
described above.  Once again, if the individual
contributions to the model-dependent
error are combined linearly, then the model-dependent errors
approximately double.
If instead the sum-rule constraint is applied to the global fit,
the same value and error for $\bar{\alpha}-\bar{\beta}$ is found.
The individual polarizabilities thus obtained are
\begin{equation}
\bar{\alpha}_{\rm global} = 12.1 \pm 0.8 \pm 0.5
\label{eq:alpha_global}
\end{equation}
and
\begin{equation}
\bar{\beta}_{\rm global} = 2.1 \mp 0.8 \mp 0.5.
\label{eq:beta_global}
\end{equation}
The error contour corresponding to the global fit with the sum-rule
constraint is shown in
Fig.~\ref{fig:ellipse}.


\section{Discussion}
\label{sec:discussion}
In recent years, various theoretical approaches have been used to calculate
the polarizabilities of the nucleon, including non-relativistic quark models,
bag models, chiral quark models, chiral perturbation theory, soliton
models, and dispersion relations.  An excellent review has been given by
L'vov \cite{L'vov:93}.  Here we elaborate on the
dispersion-relation approach, which allows
sum rules to be established for the polarizabilities.
These sum rules are completely rigorous yet
semi-phenomenological, since they relate the polarizabilities to features of
the photoabsorption cross section.  This is physically appealing, since it
helps identify the physics that gives rise to $\bar{\alpha}$ and $\bar{\beta}$,
such as the contribution of a particular nucleon resonance.

In the context of the fixed-$t$ dispersion relations that were used
to extract the polarizabilities from the Compton scattering data,
two sum rules have already been presented:  Eq.~\ref{eq:sumrules} for
$\bar{\alpha}+\bar{\beta}$ and Eq.~\ref{eq:sumruled} for
$\bar{\alpha}-\bar{\beta}$.
It is illuminating to combine the two sum rules to obtain sum rules
for $\bar{\alpha}$ and $\bar{\beta}$ separately, each
of which can be written as a sum of
an integral part and an asymptotic part:
\begin{eqnarray}
\bar{\alpha}& = & \bar{\alpha}^{\rm int} \, + \, \bar{\alpha}^{\rm asymp}
\nonumber \\
\bar{\beta}& = & \bar{\beta}^{\rm int} \, + \, \bar{\beta}^{\rm asymp}
\end{eqnarray}
The integral part can be directly evaluated using the
single-pion multipole amplitudes and the model for the multi-pion
multipole amplitudes.
The asymptotic part is just the difference between
the measured polarizability and the integral part.
The values thus obtained are
\begin{eqnarray}
\bar{\alpha}^{\rm int}\, & \approx & \, 5
\qquad
\bar{\alpha}^{\rm asymp}\,  \approx  \, 7 \nonumber \\
\bar{\beta}^{\rm int}\, & \approx & \, 8
\qquad
\bar{\beta}^{\rm asymp}\,  \approx  \, -6
\end{eqnarray}

\begin{figure}
\caption{Integrands for $\bar{\alpha}^{\rm int}$ (solid curve) and
$\bar{\beta}^{\rm int}$ (dashed curve).}
\label{fig:integrand}
\end{figure}
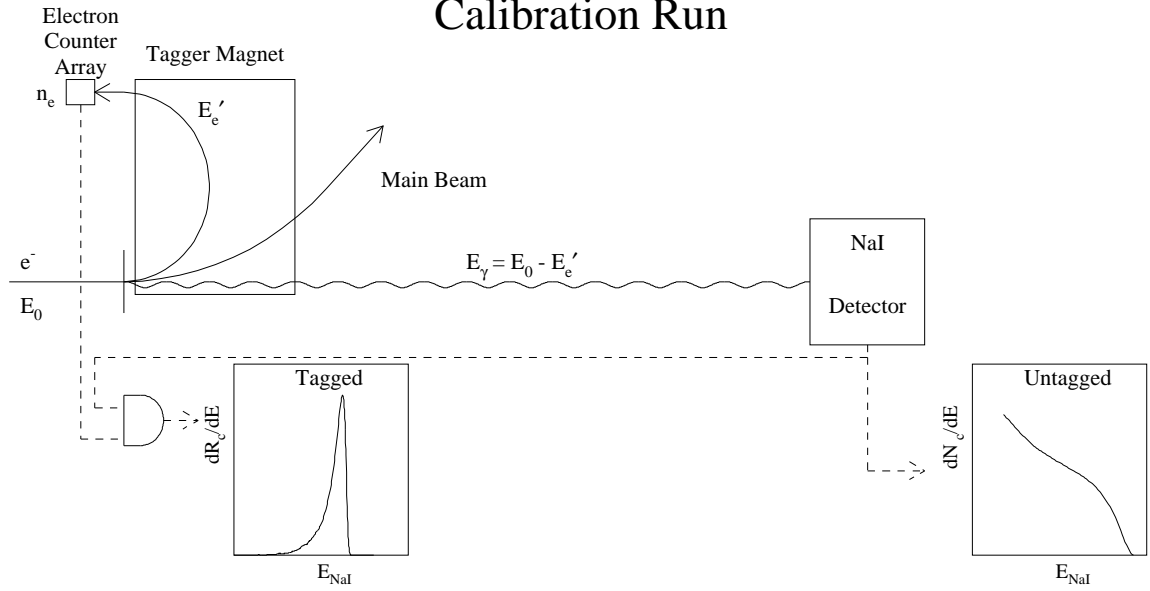
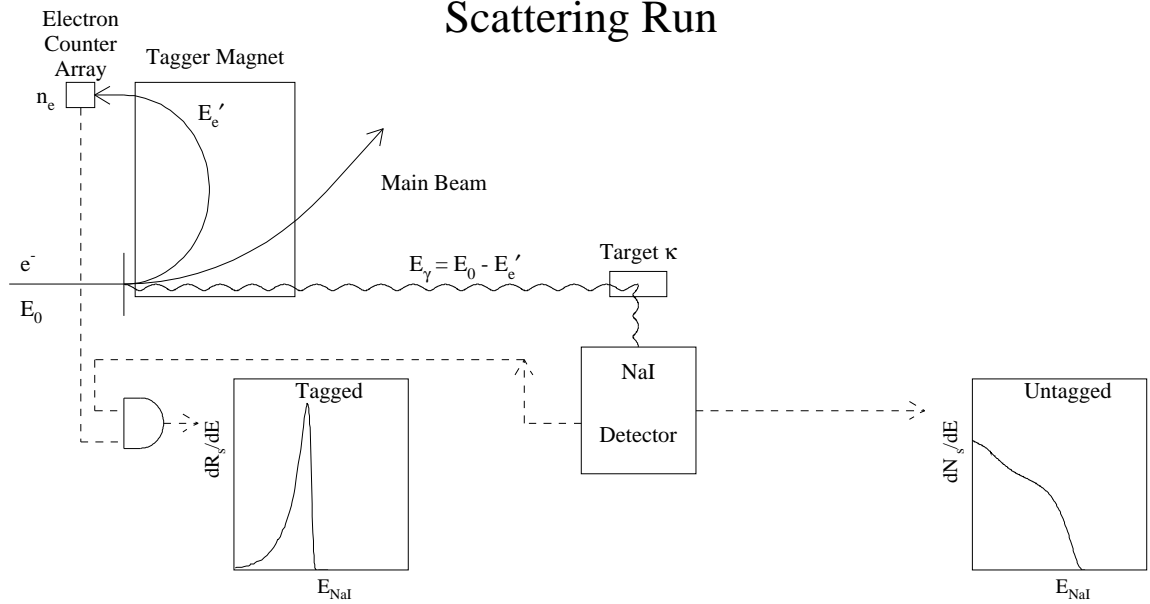
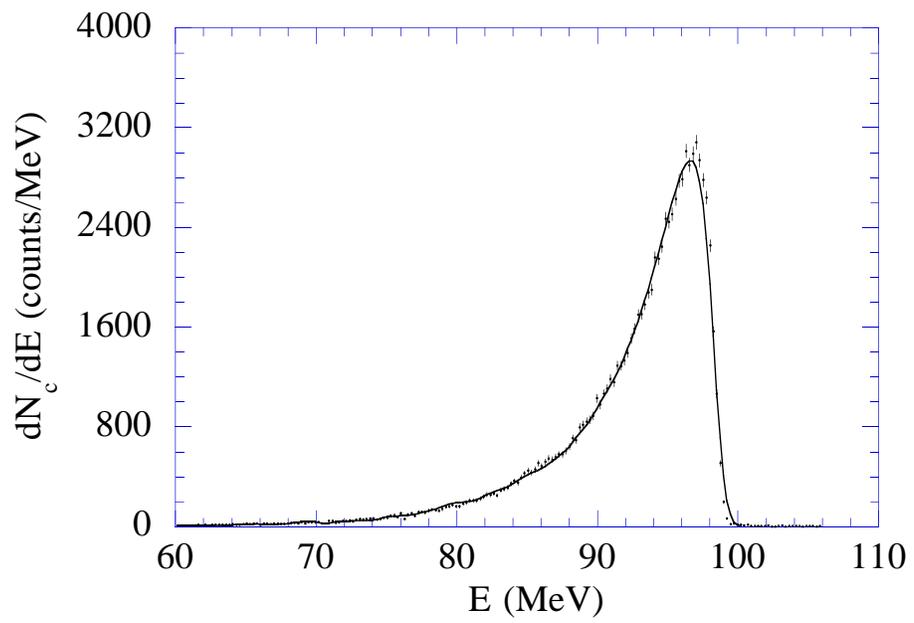
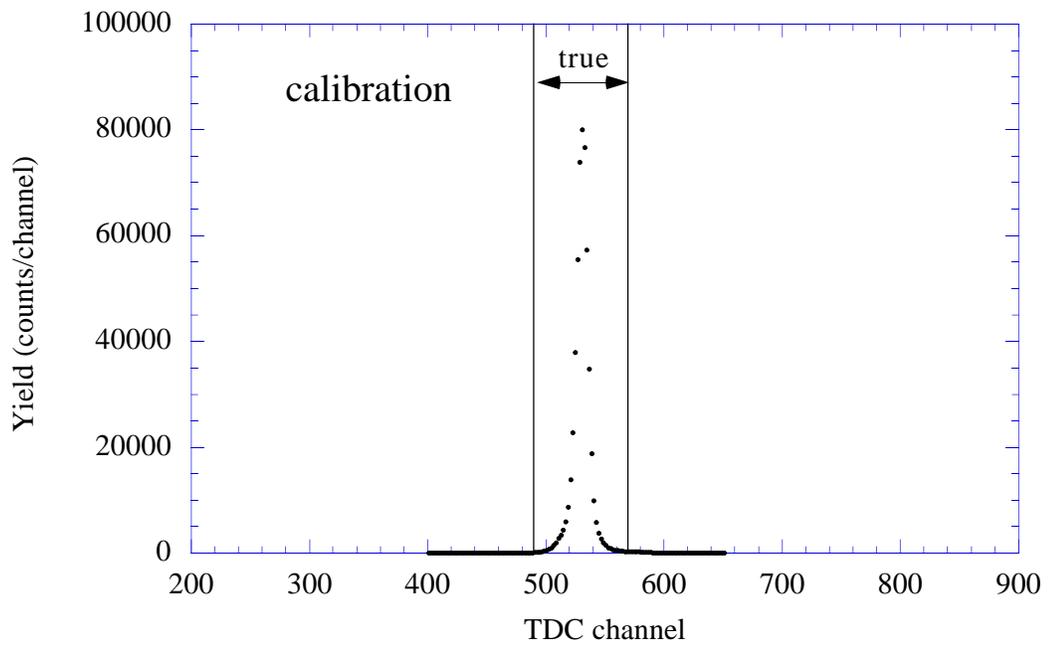
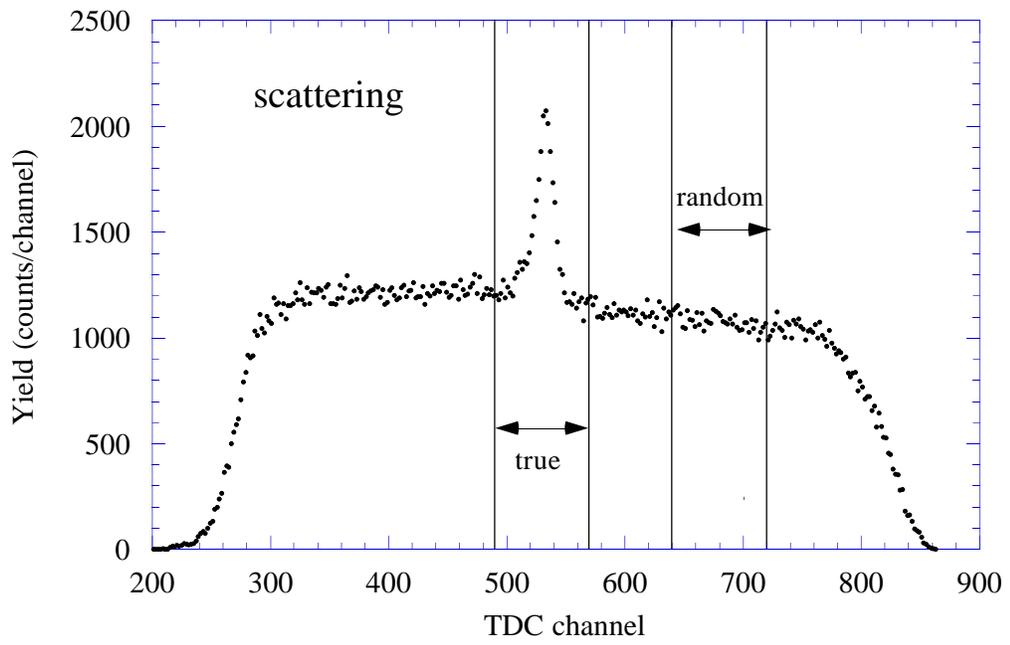
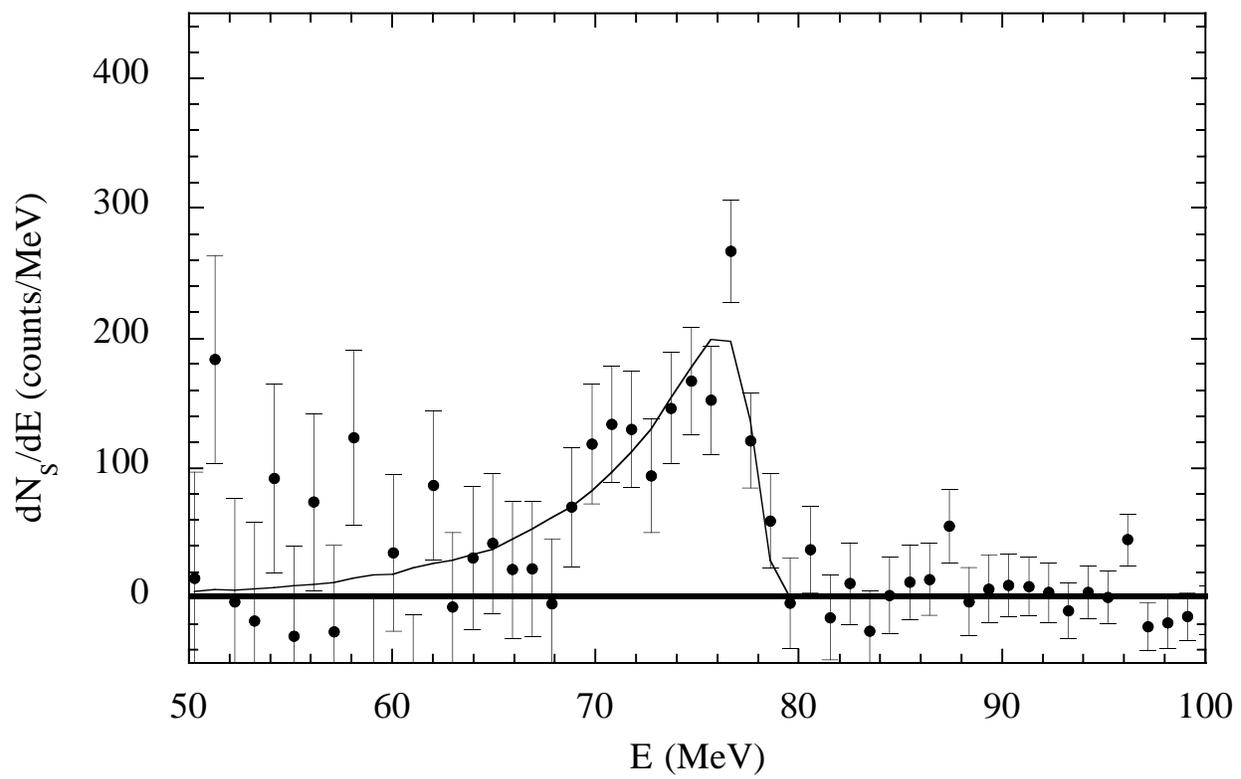
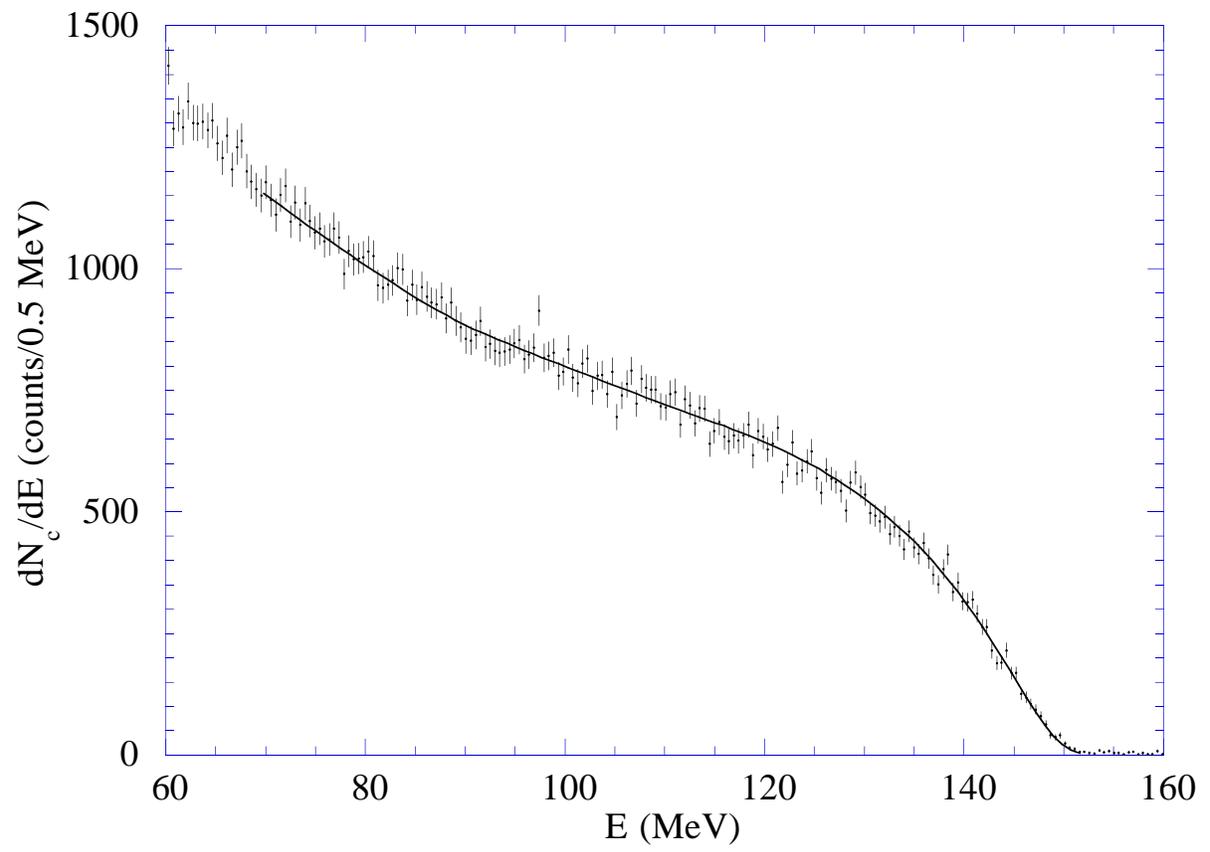
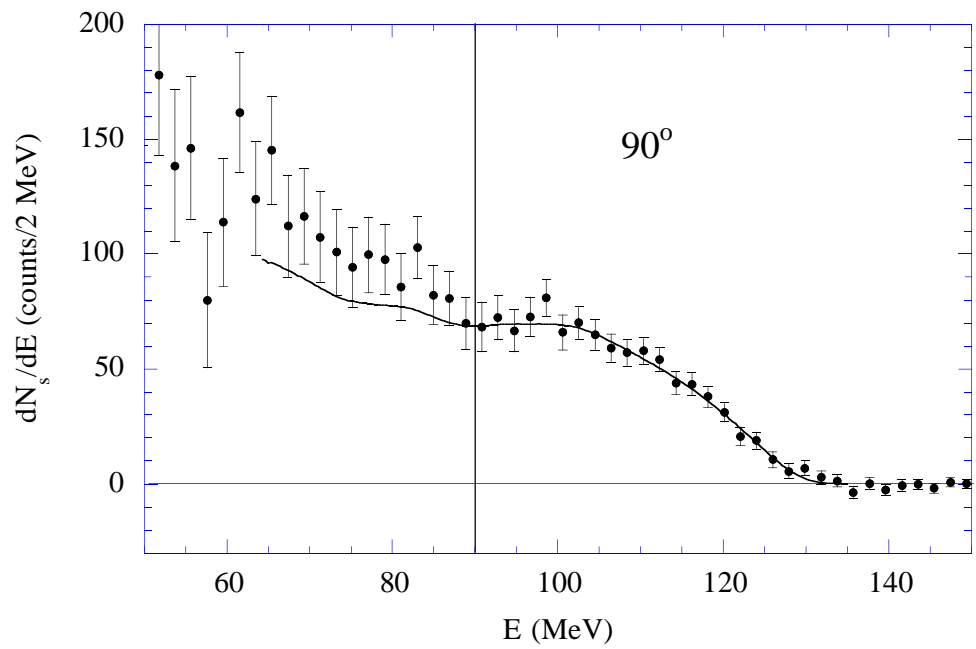
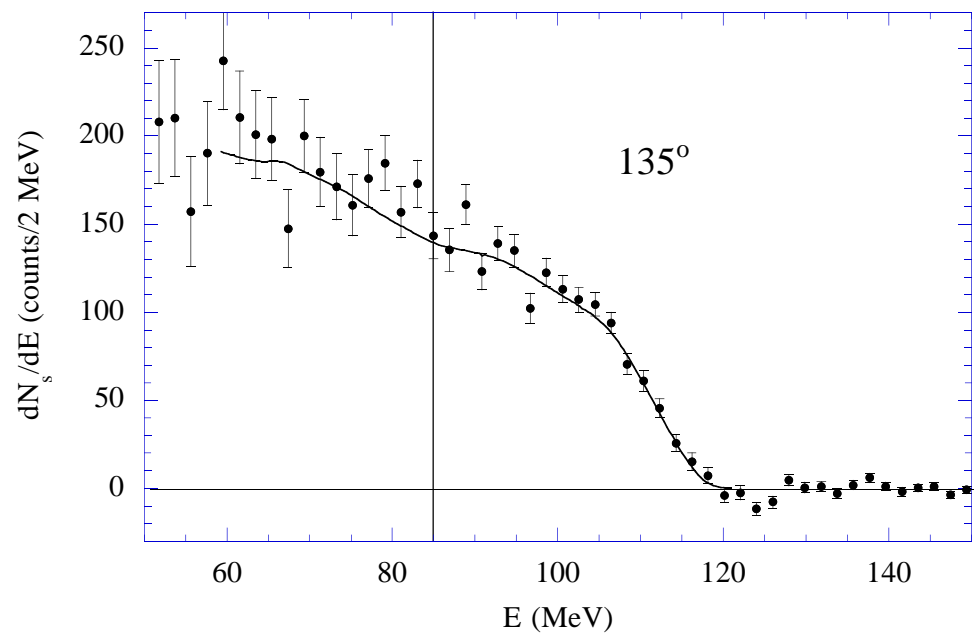
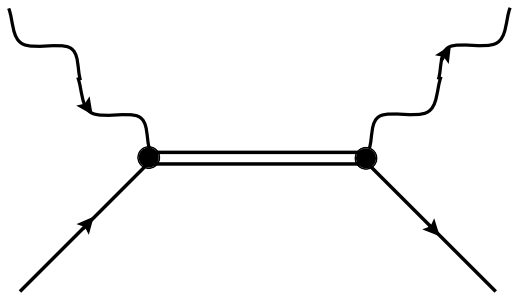
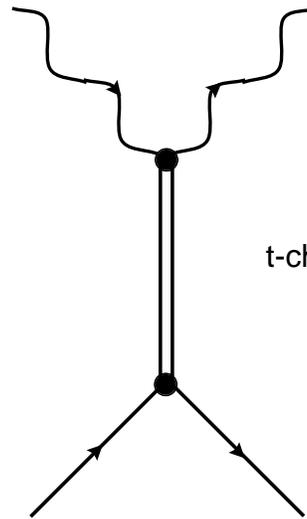
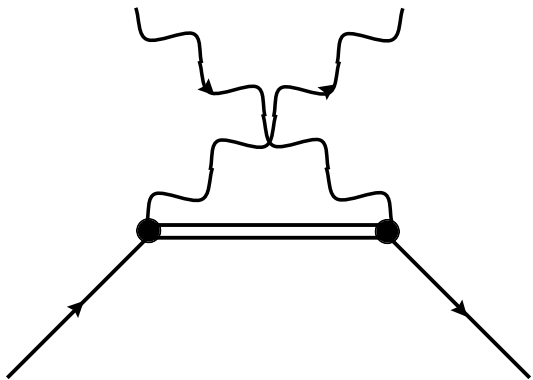
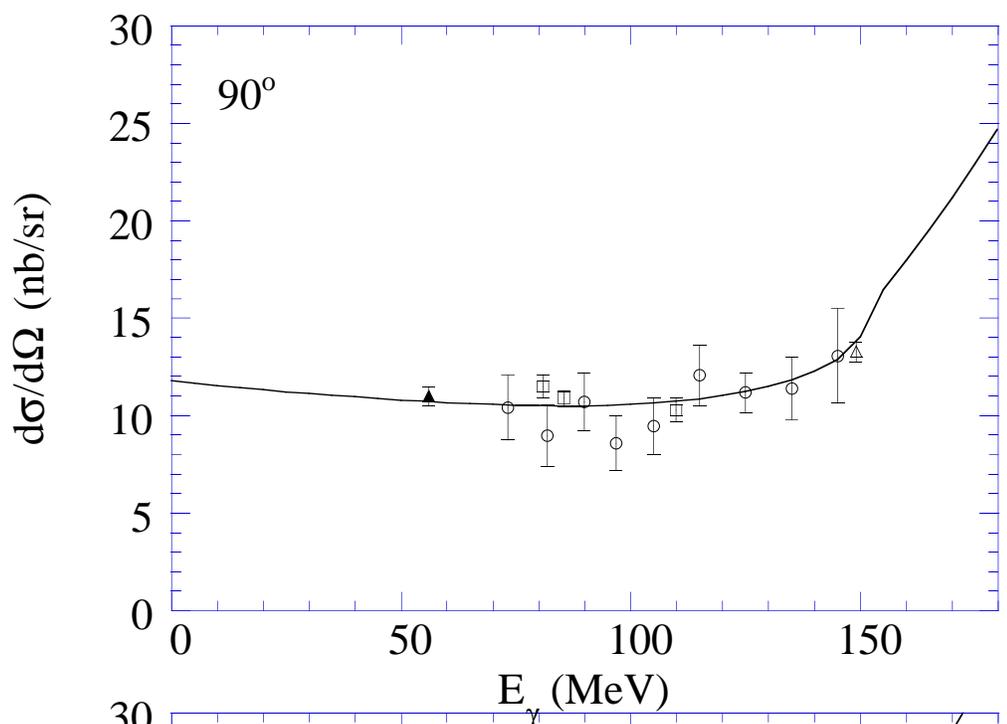
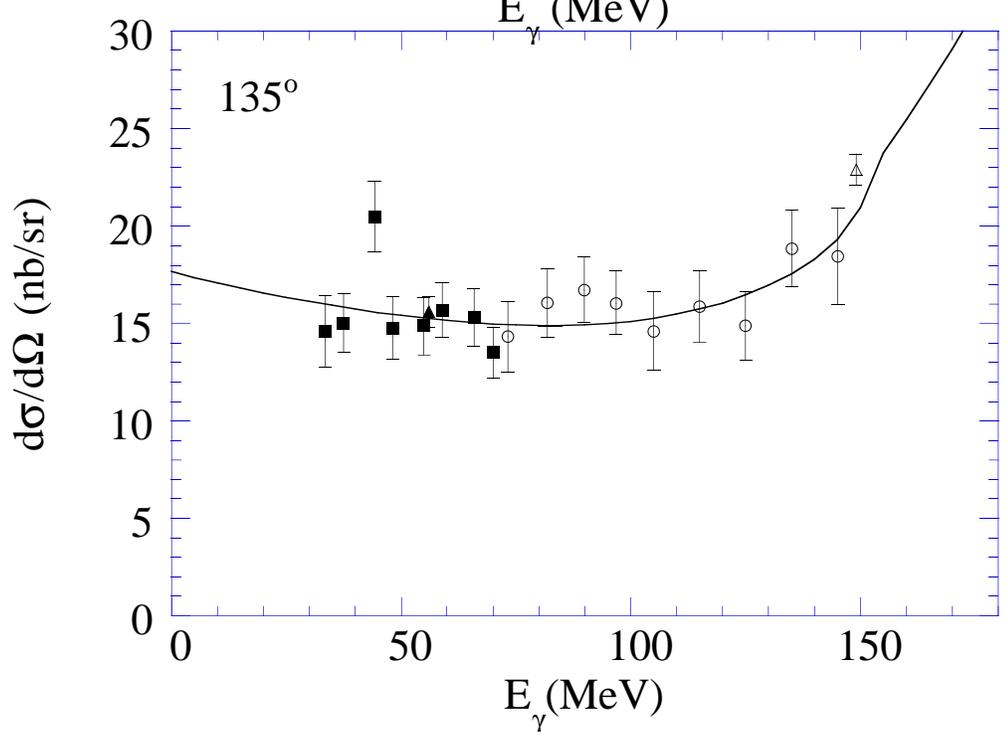
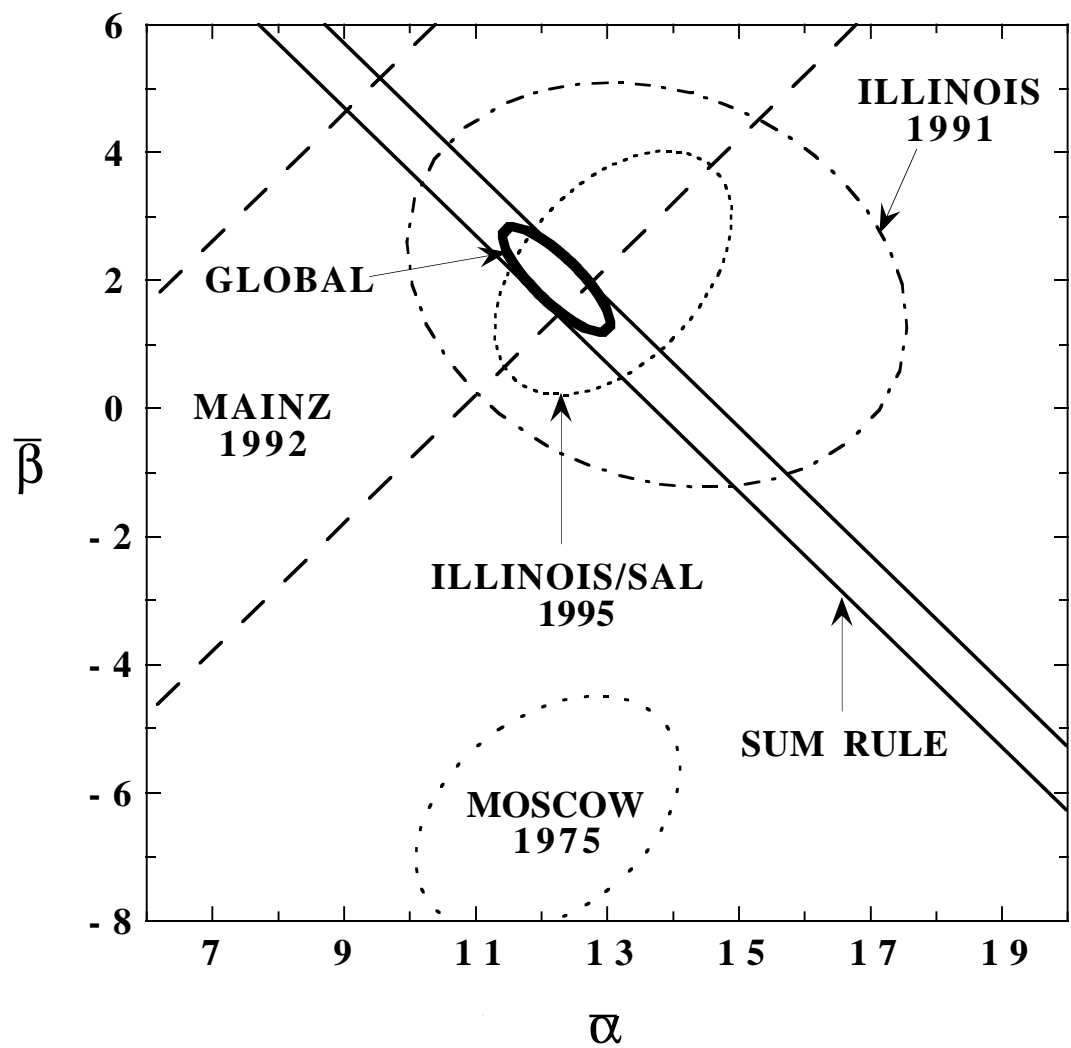
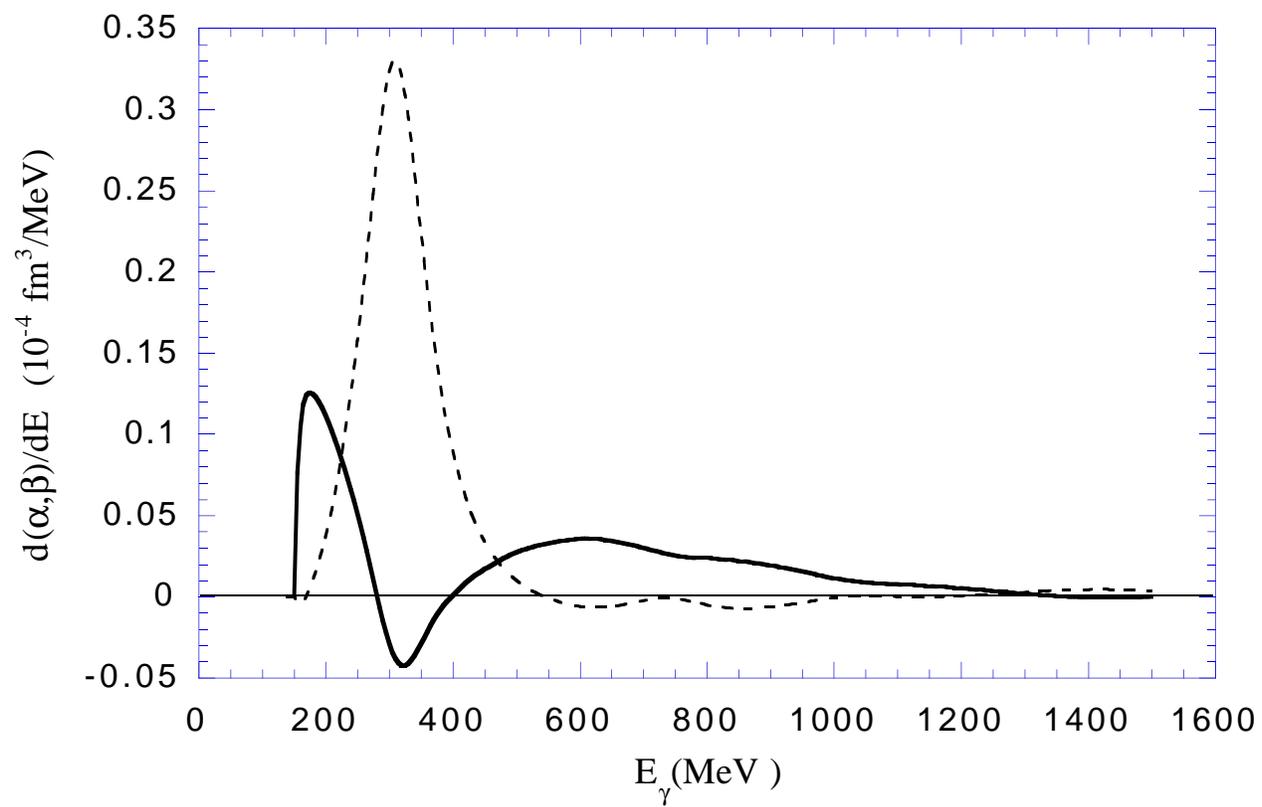

We first comment on the integral parts, the
integrands for which are shown in Fig.~\ref{fig:integrand}.
For $\bar{\alpha}^{\rm int}$ the integral is
dominated by multipoles involving non-resonant
pion photoproduction, except for a negative contribution ($\sim -3$) which
comes from the excitation of the $\Delta$ resonance.  Otherwise,
there is apparently very little contribution from degrees of freedom
associated with excitations of the valence quarks.  Indeed, a qualitative
calculation in the context of the chiral bag model \cite{Weise:85} shows
that both the electric polarizability and the diamagnetic part of the magnetic
polarizability are dominated by the polarization of the pion cloud relative
to the quark core and have little to do with the polarization of the core
itself.  This notion is confirmed by calculations using chiral perturbation
theory at the one-loop order \cite{Meissner:93}.
For $\bar{\beta}^{\rm int}$ the integral is dominated by the
$\Delta$ resonance \cite{Nimai:93}.

We next comment on the asymptotic parts of the polarizabilities, which are
neither small nor well constrained by the photoabsorption cross section.
Indeed, the combination $\bar{\alpha}-\bar{\beta}$ is almost entirely due
to the asymptotic contributions.  As discussed in
Section~\ref{sec:theoretical},
these asymptotic contributions arise primarily
from the $t$-channel exchange of a correlated pair of pions in a relative
$0^+$ state.  In effect, it arises from the scattering from the pion cloud
surrounding the nucleon.  However, as pointed out by L'vov \cite{L'vov:93},
the Born part of the scattering from the pion cloud is already largely
contained
in the integral contribution, so it is primarily the non-Born part that
enters into the asymptotic amplitude.  This idea receives support from
an alternate sum-rule approach based on a backward dispersion relation
\cite{Holstein:94}.  There is a close correspondence between the
asymptotic contribution to the fixed-$t$ sum rule and the $t$-channel
contribution to the backward sum rule, the latter involving the physical
amplitudes for the process $\gamma\gamma\rightarrow\pi\pi\rightarrow N\bar{N}$.
A substantial contribution to the $t$-channel integral is due to the non-Born
part of $\gamma\gamma\rightarrow\pi\pi$, which is essentially due to
pionic structure, including (but not limited to) the polarizability of
the pion itself.  This leads to the interesting possibility that a major
contribution to the polarizability of the nucleon is due to the internal
structure of the pions.  Since the energy needed to excite the pion is
large, this suggests that much of the physics of the nucleon polarizability
is the physics of energies beyond 1 GeV.  Any QCD-based structure model that
attempts to calculate the polarizabilities will need to address this physics.

\section{Conclusion}
\label{sec:conclusion}
The Compton scattering cross section on the proton has been measured in
the energy range 70--148 MeV, using both tagged and untagged photons.
With the aid of fixed-$t$ dispersion relations,
new values for the electric and magnetic polarizabilities have been determined
from these cross sections and the model-dependent uncertainty has been
estimated.
The present results have been combined with previously published Compton
scattering cross sections below pion threshold in order to obtain
global-average values for the polarizabilities.
Dispersion
relations indicate that the polarizabilities are only partially constrained
by existing photoproduction data and that a substantial part is due to physics
beyond 1 GeV, such as pionic structure.



\acknowledgements

It is a pleasure to acknowledge many fruitful
discussions with Dr. Anatoly L'vov and to thank him for providing us with
his dispersion code.
We gratefully acknowledge Prof. Dennis Skopik and the
staff of the Saskatchewan Accelerator Laboratory for considerable
technical and professional support during the setup and running of the
experiment.  We thank Mr. Erik Reuter for his work on the
development of the gain monitoring system and Prof. Roy Holt
for his helpful comments on this manuscript.
This research was supported in part
by the U. S. National Science Foundation under Grant Nos. NSF PHY 89-21146 and
93-10871 and by the Natural Sciences and Engineering Research Council of
Canada.


\appendix
\section*{Tables of Cross Sections}
\label{ap:cross_sections}

The differential cross sections measured in the
present experiment are tabulated in
Tables~\ref{tab:cross_tagged} and \ref{tab:cross_untagged}.  The
correlations among the untagged cross sections required the fitting
to be done in a
space where the error matrix is diagonal.  The transformation of the original
cross sections $\sigma_i=\sigma(E_i)$ (where $E_i$ = 105, 115, 125, 135, 145
MeV) from the original space to the diagonal space $\sigma_j^\prime$ has the
form
\begin{equation}
   \sigma^\prime_j = \sum_i V(j,i) \: \sigma(E_i).
\end{equation}
This transformation is written explicitly for the 135$^\circ$ detector as:
\begin{eqnarray}
\sigma^\prime_1 & = &
+0.726\:\sigma_1 -0.575\:\sigma_2 +
0.355\:\sigma_3 -0.123\:\sigma_4 +0.001\:\sigma_5  \nonumber\\
\sigma^\prime_2 & = &
+0.121\:\sigma_1 +  0.384\:\sigma_2 +
0.603\:\sigma_3 + 0.657\:\sigma_4 + 0.205\:\sigma_5 \nonumber\\
\sigma^\prime_3 & = &
-0.522\:\sigma_1 -0.245\:\sigma_2 +
0.614\:\sigma_3 -0.165\:\sigma_4 -0.513\:\sigma_5 \nonumber\\
\sigma^\prime_4 & = &
-0.334\:\sigma_1 -0.676\:\sigma_2
-0.211\:\sigma_3 + 0.580\:\sigma_4 + 0.224\:\sigma_5 \nonumber\\
\sigma^\prime_5 & = &
-0.272\:\sigma_1 -0.065\:\sigma_2 +
0.297\:\sigma_3 -0.435\:\sigma_4 + 0.803\:\sigma_5   \nonumber
\end{eqnarray}
and for the 90$^\circ$ detector as:
\begin{eqnarray}
\sigma^\prime_1 &=&  +0.206\:\sigma_1 +0.413\:\sigma_2
+0.711\:\sigma_3   +0.491\:\sigma_4   + 0.203\:\sigma_5  \nonumber\\
\sigma^\prime_2 &=& -0.611\:\sigma_1 +0.748\:\sigma_2
-0.257\:\sigma_3   +0.013\:\sigma_4   -0.036\:\sigma_5  \nonumber\\
\sigma^\prime_3 &=& -0.613\:\sigma_1 -0.293\:\sigma_2
+0.625\:\sigma_3   -0.303\:\sigma_4   -0.236\:\sigma_5  \nonumber\\
\sigma^\prime_4 &=& -0.455\:\sigma_1 -0.426\:\sigma_2
-0.181\:\sigma_3   +0.646\:\sigma_4   + 0.402\:\sigma_5  \nonumber\\
\sigma^\prime_5 &=& -0.029\:\sigma_1 +0.052\:\sigma_2
+0.078\:\sigma_3   -0.500\:\sigma_4    +0.860\:\sigma_5 . \nonumber
\end{eqnarray}
The fitting was done by transforming the calculated cross
sections at the five energies 105, 115, 125, 135, 145 MeV into the new diagonal
space, and fitting them to the uncorrelated cross sections in this space.
The transformed cross sections and their uncorrelated errors (in units of
nb/sr) for the 135$^\circ$ detector are:
\begin{eqnarray}
\sigma^\prime _1 &=&   4.461 \pm 3.009  \nonumber \\
\sigma^\prime _2 &=&  33.000 \pm 0.834 \nonumber \\
\sigma^\prime _3 &=& -14.943 \pm 2.013  \nonumber \\
\sigma^\prime _4 &=&  -3.665 \pm 1.279  \nonumber \\
\sigma^\prime _5 &=&   6.041 \pm 2.150  \nonumber
\end{eqnarray}
and for the 90$^\circ$ detector:
\begin{eqnarray}
\sigma^\prime_1 &=& 23.130   \pm 0.646 \nonumber \\
\sigma^\prime_2 &=&  0.043   \pm 1.914  \nonumber \\
\sigma^\prime_3 &=& -8.906   \pm 1.141  \nonumber \\
\sigma^\prime_4 &=&  1.145   \pm 0.977 \nonumber \\
\sigma^\prime_5 &=&  6.766   \pm 2.758.  \nonumber
\end{eqnarray}






%
%

\mediumtext

\begin{table}
\caption{Measurements of the proton polarizabilities in units
of $10^{-4}$ fm$^3$.  The values and errors
are not from the original publications but were derived by us from the
published cross sections using the techniques described in
Section~\protect\ref{sec:extraction}.  The first
error is the combined statistical
and sytematic error, and the second is an estimate of the uncertainty due to
the model.}
\begin{tabular}{lcccc}
Data Set& Energies (MeV) & Angles&
$\bar{\alpha}+\bar{\beta}$&$\bar{\alpha}-\bar{\beta}$\\ \hline
Moscow 1975 \cite{Baranov:74}	&70--110 		&90$^\circ$, 150$^\circ$
&5.8$\pm$3.3$\pm$0.2 &17.8$\pm$2.0$\pm$0.9\\
Illinois 1991 \cite{Federspiel:91}	&32--72
&60$^\circ$, 135$^\circ$
&15.8$\pm$4.5$\pm$0.1&11.9$\pm$5.3$\pm$0.2\\
Mainz 1992 \cite{Zieger:92}		&98, 132 		&180$^\circ$
&$-----$ &7.6$\pm$2.9$\pm$1.0\\
Saskatoon 1993 \cite{Hallin:93}	&149--286 		&24$^\circ$--135$^\circ$
&12.1$\pm$1.7$\pm$0.9&7.9$\pm$1.4$\pm$2.0\\
present work 	&70--148 	&90$^\circ$, 135$^\circ$
&15.0$\pm$3.1$\pm$0.4&10.8$\pm$1.8$\pm$1.0\\
\end{tabular}
\label{tab:experiments}
\end{table}

\begin{table}
\caption{Sensitivity of the extracted values of $\bar{\alpha}+\bar{\beta}$ and
$\bar{\alpha}-\bar{\beta}$ to model-dependent
uncertainties in the dispersion calculations.  The
numbers in columns 2 and 3 are the spread in values obtained from the present
experiment (in units of $10^{-4}$ fm$^3$)
 when the parameters in column 1 are changed within the range
shown or discussed in the text.  The
 last two rows show the results of combining these spreads in quadrature
and linearly.}
 \label{tab:modeldep}
 \begin{tabular}{lcc}
 Model Parameter&$\Delta(\bar{\alpha}+
\bar{\beta})$&$\Delta(\bar{\alpha}-\bar{\beta})$\\ \hline
 Single-pion multipoles&$\pm$0.15&$\pm$0.25\\
 Multi-pion multipoles&$\pm$0.25&$\pm$0.15\\
 $\pi^0$-exchange coupling ($\pm$ 4\%)&$\pm$0.00&$\pm$0.30\\
 $B_\pi$ (0--3 GeV$^{-2}$)&$\pm$0.00&$\pm$0.65\\
 $B_{2\pi}$ (6--10 GeV$^{-2}$)&$\pm$0.25&$\pm$0.60\\
 $B_P$ (6--10 GeV$^{-2}$)&$\pm$0.00&$\pm$0.00\\ \hline
 combined in quadrature&$\pm$0.38&$\pm$0.98\\
 combined linearly&$\pm$0.65&$\pm$1.95\\
 \end{tabular}
 \end{table}

\begin{table}
\caption{Tagged differential cross sections for
each angle and the statistical error.}
\begin{tabular}{|c|c|c|}
  Central Photon Energy (MeV) & $\frac{d\sigma}{d\Omega}(135^\circ)$ (nb/sr)&
                   $\frac{d\sigma}{d\Omega}(90^\circ)$ (nb/sr) \\ \hline
73.2 & $14.33 \pm 1.81$ & $10.41 \pm 1.66$ \\
81.8 & $16.06 \pm 1.77$ & $ 8.97 \pm 1.60$ \\
89.8 & $16.71 \pm 1.68$ & $10.72 \pm 1.49$ \\
96.8 & $16.05 \pm 1.63$ & $ 8.61 \pm 1.40$ \\
\end{tabular}
\label{tab:cross_tagged}
\end{table}

\begin{table}
\caption{Untagged differential cross sections
for each angle along with the corresponding
diagonal terms in the error matrix.}
\begin{tabular}{|c|c|c|}
Energy Range (MeV) & $\frac{d\sigma}{d\Omega}$(135$^\circ$)(nb/sr) &
$\frac{d\sigma}{d\Omega}$(90$^\circ$)(nb/sr) \\ \hline
100--110 & 14.60  $\pm$ 2.53 & 9.48   $\pm$ 1.44\\
110--120 & 15.87  $\pm$ 2.03 & 12.07  $\pm$ 1.56\\
120--130 & 14.88  $\pm$ 1.84 & 11.19  $\pm$ 1.02\\
130--140 & 18.85  $\pm$ 1.40 & 11.40  $\pm$ 1.59\\
140--150 & 18.45  $\pm$ 2.04 & 13.07  $\pm$ 2.42\\
\end{tabular}
\label{tab:cross_untagged}
\end{table}
%
%
\newpage
\begin{figure}
\caption{Calculations of the Compton scattering cross section from the proton,
showing the cross section for the point proton (Born), the LEX, and
the fixed-$t$ dispersion relations (DR).  The calculations assume that
$\bar{\alpha}+\bar{\beta}=14.2$ and $\bar{\alpha}-\bar{\beta}=9.0$.}
\label{fig:cross_section_plot}
\end{figure}

\begin{figure}
\caption{A schematic of the photon scattering techniques used in this
experiment, showing the calibration measurement (upper half)
and the scattering measurement (lower half).  Sample tagged and untagged
photon spectra were generated by a Monte Carlo simulation.}
\label{fig:photon}
\end{figure}

\begin{figure}
\caption{Tagged photon calibration spectrum and Monte Carlo fit.}
\label{fig:tagged_brems}
\end{figure}

\begin{figure}
\caption{TDC spectra for the 135$^\circ$ detector, summed over all channels
of the tagger.  The TDC is started by a photon signal from the NaI
and stopped by an electron signal from the tagger focal plane.
Each TDC channel corresponds to 0.1 ns.  The cuts applied for the
true and random coincidences are indicated.}
\label{fig:tdc}
\end{figure}

\begin{figure}
\caption{Tagged spectrum of photons scattered from hydrogen at 135$^\circ$
and Monte Carlo fit.}
\label{fig:tagged_H}
\end{figure}

\begin{figure}
\caption{Untagged photon calibration spectrum and Monte Carlo fit.}
\label{fig:untagged_brems}
\end{figure}

\begin{figure}
\caption{Untagged spectra of photons scattered from hydrogen at
90$^\circ$ and 135$^\circ$ and
Monte Carlo fits.
Using the tagged cross sections, the calculation has been extended into the
tagging region, which is to the left of the vertical
line.}
\label{fig:untagged_H}
\end{figure}

\begin{figure}
\caption{Diagrams for Compton scattering from the proton, including
$s$-channel and $u$-channel processes as well as
$t$-channel exchanges.}
\label{fig:diagrams}
\end{figure}

\begin{figure}
\caption{Cross sections measured in the present experiment (open circles)
 and curves
calculated with the fixed-$t$ dispersion relations
using
$\bar{\alpha}+\bar{\beta}$=14.2, $\bar{\alpha}-\bar{\beta}$=10.8, and a
particular set of model parameters.
Also shown are the data of Goldansky (closed triangles) \protect\cite{Go60},
Baranov (open squares)
\protect\cite{Baranov:74},
Federspiel (closed squares) \protect\cite{Federspiel:91}, and Hallin
(open triangles) \protect\cite{Hallin:93}.}
\label{fig:fit}
\end{figure}

\begin{figure}
\caption{Error contours in the $\bar{\alpha}\bar{\beta}$-plane
for the experiments listed in
Table~\protect\ref{tab:experiments} and for the dispersion sum rule.
Also shown is the error contour for a global fit to all the data, excluding
the Moscow 1975 cross sections, as described in the text.
The contours correspond to one standard deviation for the combined
statistical and systematic errors, and the values are for one set of
model parameters.}
\label{fig:ellipse}
\end{figure}

\end{document}